\documentclass[%
 reprint,
 twocolumn,
 amsmath,amssymb,
aip,jcp,
]{revtex4-2}

\usepackage{natbib}
\usepackage{graphicx}
\usepackage{dcolumn}
\usepackage{bm}
\usepackage{xcolor}
\usepackage[version=3]{mhchem}

\begin{document}
\author{Daniel Weinberg}
\email{d\textunderscore weinberg@berkeley.edu}
\address{Department of Chemistry, University of California, Berkeley, California 94720, USA\looseness=-1}
\address{Materials Sciences Division, Lawrence Berkeley National Laboratory, Berkeley, California 94720, USA \looseness=-1}

\author{Yoonjae Park}%
\address{Department of Chemistry, University of California, Berkeley, California 94720, USA\looseness=-1}
 \address{Materials Sciences Division, Lawrence Berkeley National Laboratory, Berkeley, California 94720, USA\looseness=-1}

\author{David T. Limmer}
 \address{Department of Chemistry, University of California, Berkeley, California 94720, USA\looseness=-1}
 \address{Materials Sciences Division, Lawrence Berkeley National Laboratory, Berkeley, California 94720, USA\looseness=-1}
 \address{Chemical Sciences Division, Lawrence Berkeley National Laboratory, Berkeley, California 94720, USA\looseness=-1}
 \address{Kavli Energy NanoScience Institute, Berkeley, California 94720, USA\looseness=-1}
\author{Eran Rabani}
\email{eran.rabani@berkeley.edu}
\address{Department of Chemistry, University of California, Berkeley, California 94720, USA\looseness=-1}
\address{Materials Sciences Division, Lawrence Berkeley National Laboratory, Berkeley, California 94720, USA\looseness=-1}
\address{The Raymond and Beverly Sackler Center of Computational Molecular and Materials Science, Tel Aviv University,
Tel Aviv 69978, Israel\looseness=-1}

\title{Size-dependent lattice symmetry breaking determines the exciton fine structure of perovskite nanocrystals}

\begin{abstract}
The ordering of optically bright and dark excitonic states in lead-halide perovskite nanocrystals has been a matter of some debate. It has been proposed that the unusually short radiative lifetimes in these materials is due to an optically bright excitonic ground state, a unique situation among all nanomaterials. This proposal was based on the influence of the Rashba effect driven by lattice-induced inversion symmetry breaking. Direct measurement of the excitonic emission under magnetic fields has shown the signature of a dark ground state, bringing the role of the Rashba effect into question. Here, we use a fully atomistic theory to model the exciton fine structure of perovskite nanocrystals accounting for the realistic lattice distortion at the nanoscale. We calculate optical gaps and exciton fine structure that compare favorably with a wide range of experimental works. We find a non-monotonic dependence of the exciton fine structure splittings due to a size dependence structural transition between cubic and orthorhombic phases. In addition, the excitonic ground state is found to be dark with nearly pure spin triplet character resulting from a small Rashba coupling. We additionally explore the intertwined effects of lattice distortion and nanocrystal shape on the fine structure splittings, clarifying observations on poly-disperse nanocrystals. 
\end{abstract}

\maketitle

Lead-halide perovskite nanocrystals (NCs) have attracted significant attention due to their remarkable optical and electronic properties that could lend themselves to diverse applications.\cite{Kovalenko2017, Protesescu2015, Schmidt2014} 
Perhaps most interestingly, these materials show remarkably fast radiative lifetimes, which shorten at low temperatures in contrast to other nanomaterials~\cite{Canneson2017,Becker2018,Chen2018,Tamarat2019, Xu2019,Rossi2020-NanoLet,Rossi2020-JCP}. This anomalous temperature dependence of the radiative lifetimes has led to speculation that these materials could exhibit a reversal of the typical exciton fine structure (FS) measured in all other nanomaterials to date~\cite{Nirmal1995,Becker2018,Sercel2019}. Specifically, Becker et al.~\cite{Becker2018} proposed that the lowest excitonic state is a bright state, i.e that it has an optically allowed transition to the material ground state.

If that is the case, at low temperatures the carriers will preferentially be in the bright, rapidly emissive state rather than depending on thermal fluctuations to reach an emissive state. Understanding the excitonic fine structure in these materials is important to assess their suitability as quantum light sources, which depends in part on the uniquely fast radiative lifetimes.\cite{Aharonovich2016,Park2015,Utzat2019,Lv2019} The argument for a bright excitonic ground state was supported by a detailed analysis of the physics of excitons in perovskite NCs from an effective mass model. We will briefly revisit this before describing our atomistic approach to this problem, which can provide a definitive ordering of bright and dark states in lead-halide perovskite NCs.

In the typical picture, electrons and holes are bound into excitons by their strong Coloumbic attraction forming a hydrogenic series of states that may be modified by confinement effects of the NC.\cite{Scholes2006} For systems with negligible spin-orbit coupling, the electron and hole spins are decoupled from the spatial degrees of freedom and simple addition of angular momentum describes the resulting triplet and singlet spin functions. The electron-hole exchange interaction slightly reduces the strength of exciton binding for excitons with spin-singlet character, introducing a spin dependence into the exciton FS. Notably, for materials like perovskites with significant spin-orbit coupling, the spatial and spin degrees of freedom are not separable. Further, it is known that the excited state properties of the perovskites are sensitive to the lattice structure, as the charge-lattice coupling in these materials is significant.\cite{park2022nonlocal,park2022renormalization,schilcher2021significance,Mayers2018}

In the specific case of the perovskite materials, the conduction band is composed mainly of of Pb-6p orbitals which are strongly split by spin-orbit coupling. The conduction band edge is composed of the the $J=1/2$ total angular momentum subspace formed from the addition of the spin and the orbital angular momentum. The valence band has s-type symmetry and thus is not split by spin-orbit coupling.\cite{Brivio2014} In the exciton, this causes the exchange interaction to split three bright states above a dark ground state, all with mixed spin-triplet and spin-singlet character. These bright states each have dipoles polarized along one of the principal axes, and for a cubic crystalline structure are perfectly degenerate. These three bright states are often referred to as ``bright triplets" due to their total angular momentum triplet character, however this should not be confused with their spin character. 

\begin{figure}
    \centering
    \includegraphics[width=240pt]{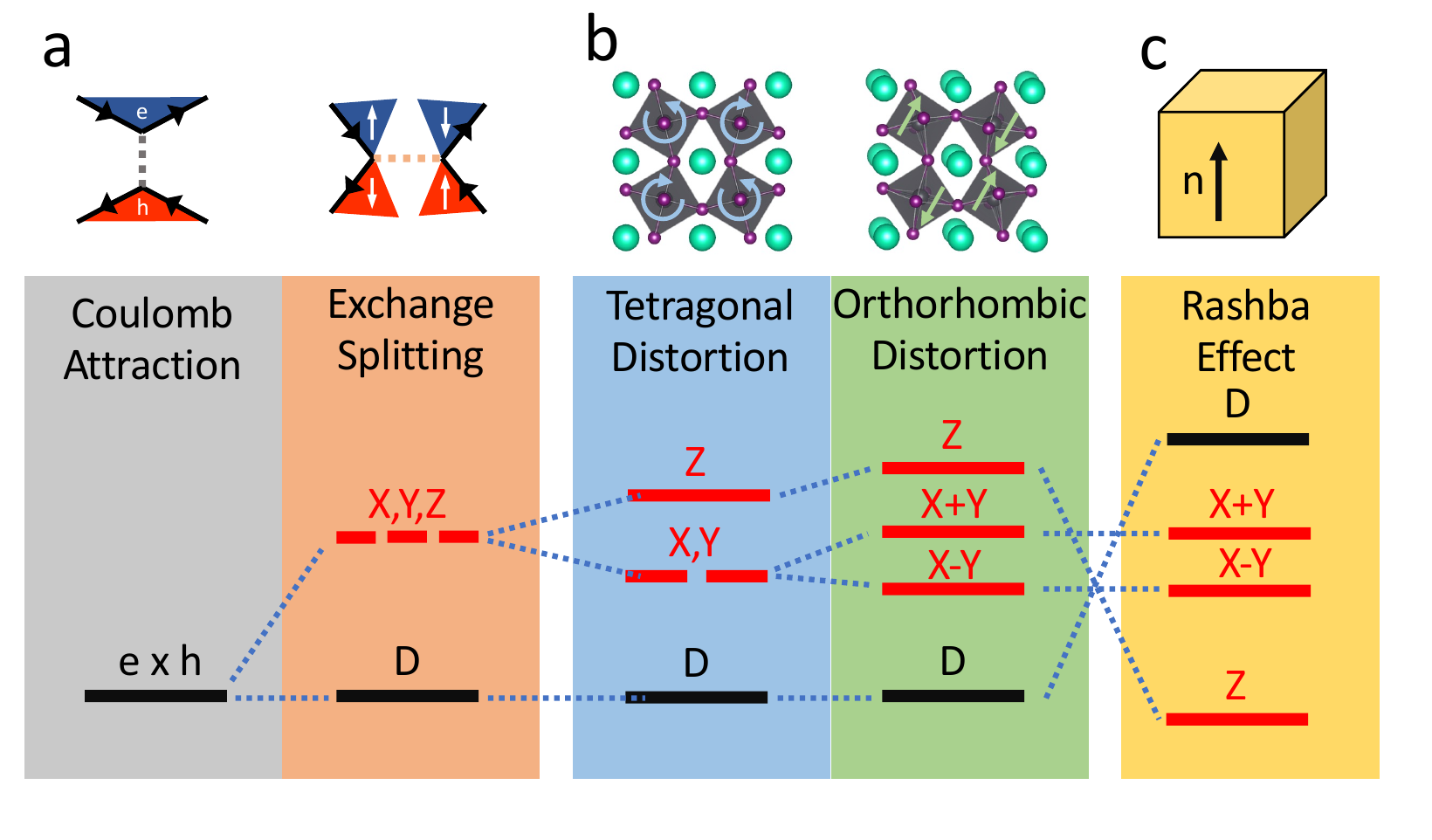}
    \caption{(a) The effects of electron-hole interaction, (b) lattice distortion, and (c) the previously proposed\cite{Becker2018} role of the Rashba effect on the exciton FS.}
    \label{fig:figScheme}
\end{figure}
    
Any deviation from this cubic structure will result in splitting among these bright states. Perovskite materials are known to progress through a series of symmetry lowering phase transitions as temperature is reduced,\cite{Dastidar2017,Quarti2016} and Figure \ref{fig:figScheme}(b) illustrates the effect of these distortions on the exciton FS. A tetragonal distortion caused by rotation of the lead-halide octahedra around the z-axis splits the z-oriented state higher in energy, but the symmetry in the x-y plane maintains the degeneracy of those two states. At lower temperatures, tilting of the octahedra breaks that symmetry and further splits the bright states in the x-y plane into two states polarized at 45$^\circ$ angle to the principle axes~\cite{Han2022}. These symmetry lowering splittings are observed experimentally as a splitting of the excitonic emission into two or three distinct lines~\cite{Becker2018,Tamarat2019,Tamarat2020}, but on their own they do not lead to a reversal of the bright-dark level ordering. 

The theorized bright ground state may be arrived at through the influence of the Rashba effect.\cite{Rashba1959} This additional term in the $\mathbf{k}\cdot \mathbf{p}$ Hamiltonian comes from the co-existence of strong spin-orbit coupling and inversion symmetry breaking. The additional ``effective exchange"\cite{Becker2018} term only enters the Hamiltonian through two parameters-- a magnitude and direction along which inversion symmetry is broken. This, by nature, is blind to the atomistic detail of the symmetry breaking at the nanoscale, and leaves unknown the exact nature of a NC structure that would give rise to such a level ordering. 

 In fact, several recent measurements~\cite{Tamarat2019,Tamarat2020,Hou2021} have detected a signature of a dark ground state several meVs below the bright states. Under the influence of a external magnetic field, the Zeeman effect  couples dark states to energetically close bright states, resulting in an emergent emission line. Under these conditions the dark ground state can be directly observed. The fine structure splittings are instead explained in terms of the interplay of crystal structure and NC shape anisotropy.\cite{Tamarat2020} 

Various theoretical attempts have been made to provide additional understanding of atomistic detail of this effect, as well as provide tools to understand how to disentangle the Rashba effect, the crystal field splitting, and NC shape anisotropy to determine the level ordering and splitting in these NCs. Within an effective mass model, the Rashba splitting as indicated by the energy difference between Z and X/Y excitons is predicted to increase to the bulk limit with increasing NC size.\cite{Sercel2019} On the other hand, the effect of shape anisotropy should be lesser for larger NCs.\cite{Sercel2019-2} This does, however, lead to a troubling question: If the Rashba effect is more pronounced for larger NCs, but absent in the bulk, where would the transition to more bulk-like behavior occur? The resolution of this must come from an atomistic theory that can also describe how the structure of small NCs may be distorted and how that of large NCs converges to the bulk limit. A recent theoretical investigation focusing on methylammonium lead iodide considered the effect of methylammonium relaxation within a fixed tetragonal lead iodide framework and found only weak a Rashba effect insufficient to cause level inversion.\cite{Biffi2022}

\begin{figure*}
    \centering
    \includegraphics[width=5in]{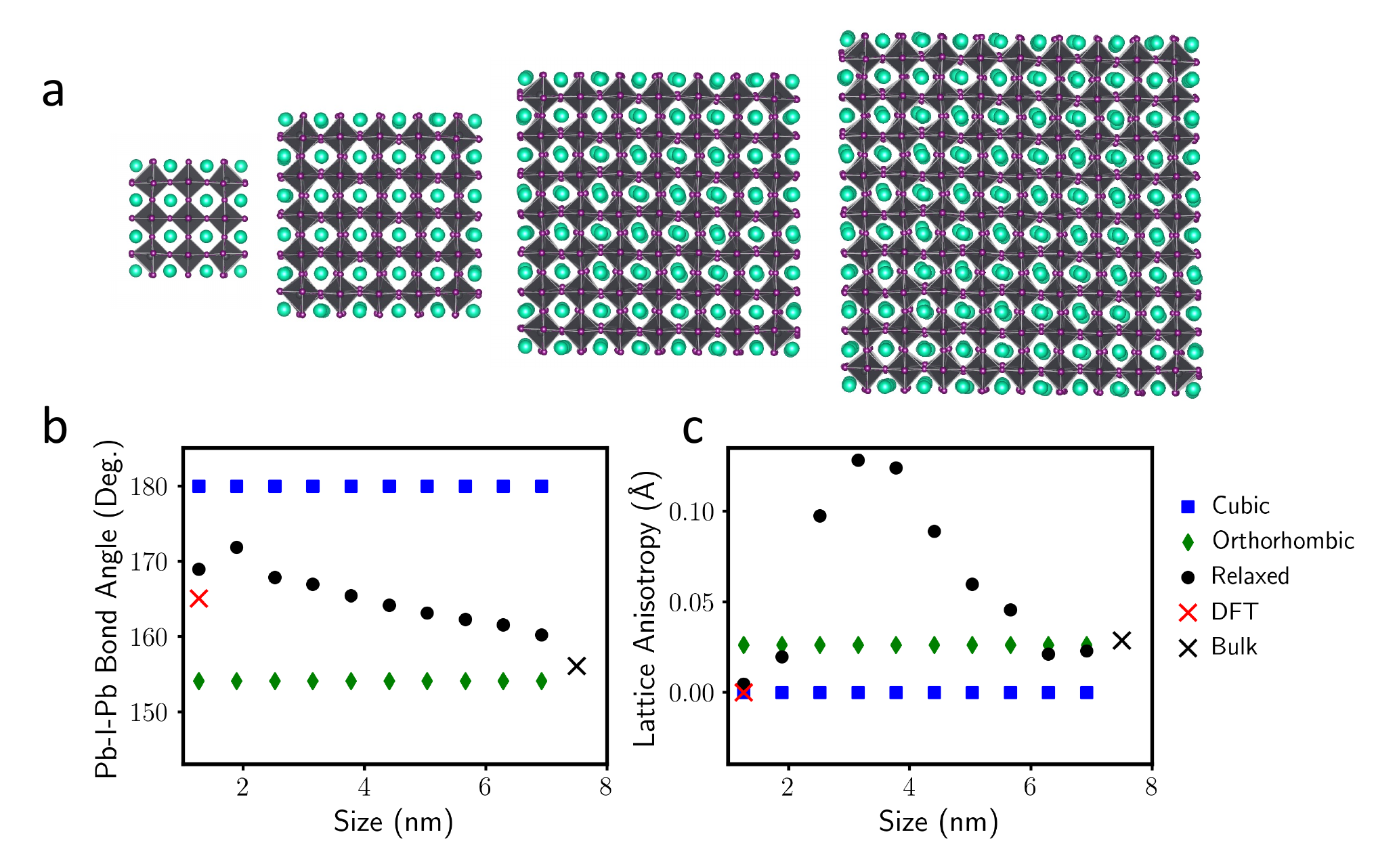}
    \caption{(a) Renderings of 1.9nm, 3.1nm, 4.4nm, and 5.7nm \ce{CsPbI_3} NC cubes after structural relaxation. Cs atoms are shown in teal, I atoms in purple and Pb are shown as grey coordination octahedra. (b) The average Pb-I-Pb bond angle and (c) The extent of lattice anisotropy induced by the relaxation for cubic (blue squares), orthorhombic (green diamonds) and relaxed (black circles) structures.  }
    \label{fig:struct}
\end{figure*}

To fully understand the intertwined roles of the Rashba effect and lattice symmetry we must consider the full structure of perovskite NCs in atomistic detail, especially including the lead halide framework that contributes most strongly to the valence and conduction band states. To do this, we use a previously developed atomistic force field\cite{Bischak2020} to find the lowest energy configuration for a series of \ce{CsPbI_3} perovskite NCs shown in Figure \ref{fig:struct}(a). The bulk properties of this model have been extensively validated.\cite{quan2021vibrational,limmer2020photoinduced} As the measurements of the exctionic FS occur at cryogenic temperatures, these single minimized structures accurately represent the atomic configuration of the NCs in these experiments, and the effects of lattice dynamics may be ignored. The relaxed structures can be compared to the bulk cubic and orthorhombic structures on the basis of the average Pb-I-Pb bond angles. These are shown in Figure \ref{fig:struct} (b) and reveal that these relaxed structures lie somewhere between the cubic and orthorhombic structures. The cubic structures have no octahedral rotation and therefore all bond angles are 180 degrees. For the orthorhombic structures the significant octahedral rotation leads to an average bond angle of 154 degrees. The smallest relaxed structures take more cubic forms, but the larger ones approach the orthorhombic configuration which is the stable bulk structure.

 To quantify the extent to which the NC relaxation breaks crystal symmetries we define a lattice anisotropy parameter. It is defined by taking the average of the Pb-Pb distances along each of the principal axes, and then finding the difference between the direction with the lowest average and the direction with the highest average. We plot this parameter against NC size in Figure \ref{fig:struct} (c). For the cubic structures this is always zero, and for the orthorhombic structures the elongated z-axis gives a small constant anisotropy. The small relaxed structures are highly symmetric so this anisotropy is near zero, but as the size increases beyond 2~nm in size, octahedral rotations begin to emerge. These are not uniform throughout the NC, however, and remain suppressed at the surface leading to significant lattice anisotropy. Significant deviation from the cubic crystal structure is not unexpected, as although cubic phase QDs have been stabilized at room temperature and somewhat below,\cite{Swarnkar2016} the cryogenic temperatures at which the FS measurements take place should favor the orthorhombic structure. The size-dependent effect has been observed experimentally\cite{Zhao2020}, and is driven by a competition between surface energy and bulk phase stability which had been previously explored using a continuum model\cite{Yang2020}. The predicted phase crossover around 2.7~nm aligns well with the region of highest lattice anisotropy. This lends confidence to our ability to produce an atomistic description of complex structural behavior at the nanoscale. This size dependent effect has not previously been considered in the context of the exciton FS, and will play a crucial role in understanding the size dependence of the FS splittings. 

Obtaining the exciton FS from these relaxed NC configurations requires an electronic structure method that is responsive to the atomistic detail of the material. While these materials are too large for \textit{ab-initio} theories such as DFT combined with many-body perturbation techniques, semi-empirical methods are able to access the size ranges necessary. We employ the semi-empirical pseudopotential method~\cite{Wang1994,Wang1996,Rabani1999b,Williamson2000}, which assigns each atom in the NC an effective potential derived from bulk band structures. These pseudopotentials include both local and non-local components that capture the effect of spin-orbit coupling. As our relaxed NCs lie somewhere in between the cubic and orthorhombic crystal phases, pseudopotentials have been fit to describe the band structures of both phases individually. The pseudopotentials used in the NC calculations are linearly interpolated between these, based on the local NC structure. This way, the electronic structure is sensitive to local deformations or distortions in the lattice. The optical absorption spectrum is computed using the Bethe-Salpeter equation (BSE) within the static screening approximation, which describes the bound excitonic states in the basis of free electron-hole pairs.\cite{Rohlfing2000,Eshet2013} This approach allows for equal treatment of the direct and exchange terms in a non-perturbative manner, and fully takes into account the effects of spin-orbit coupling. This treatment is essential to determining the full excitonic spectrum of these NCs and the FS splitting. Additional details on the electronic calculations can be found in the Supporting Information.

\begin{figure*}
    \centering
    \includegraphics[width=15cm]{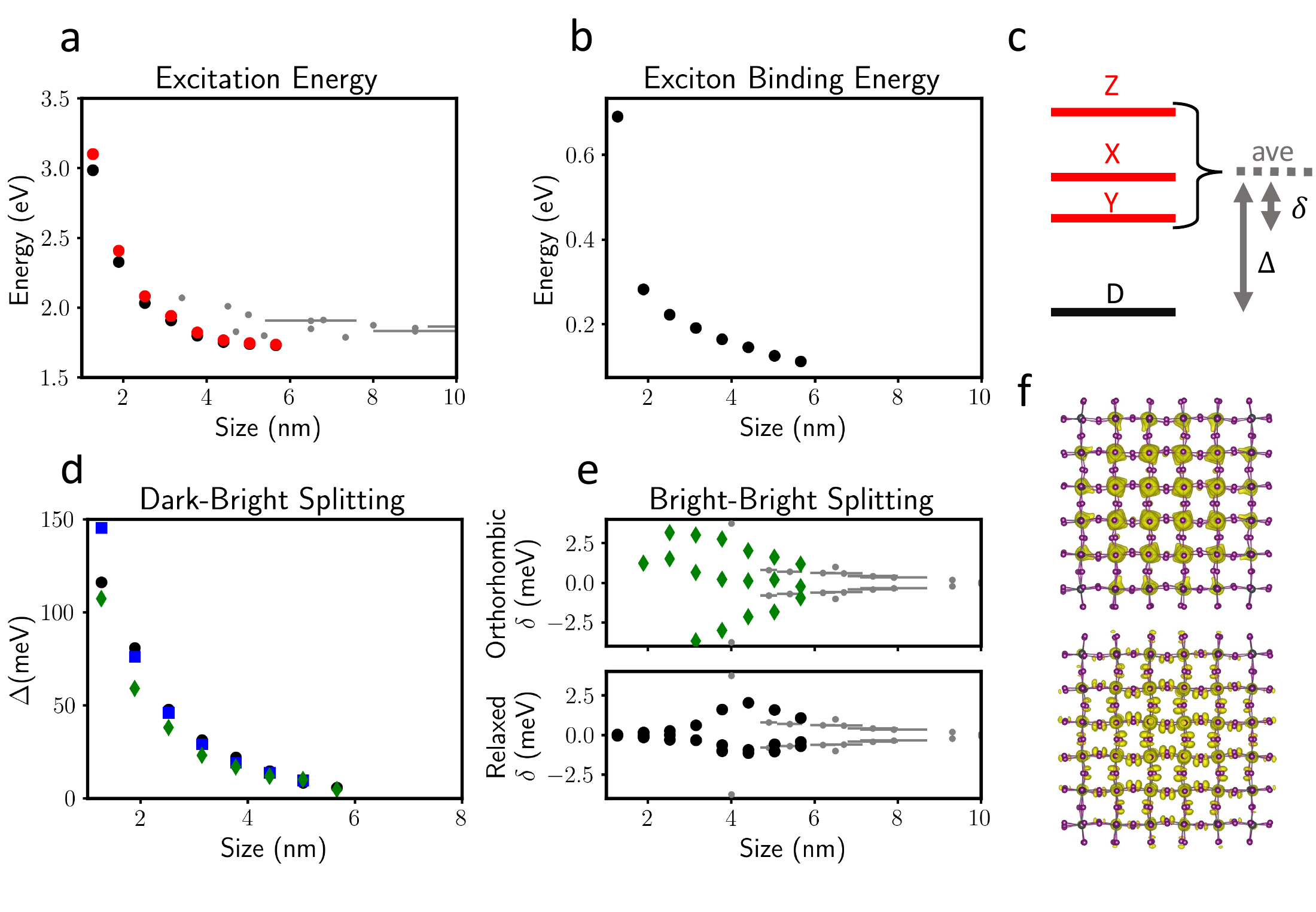}
    \caption{(a) The lowest bright (red) and dark (black) excitonic states for relaxed NCs as a function of size along with experimental data \cite{Qiao2021,Shang2019,Yao2019,Guvenc2019, Swarnkar2016, Paul2022}. (b) Exciton binding energies for relaxed NCs as a function of size. (c) A level diagram describing the splittings calculated. (d) Calculated splitting between the bright and dark excitonic states. (e) Splitting among the bright excitonic states for orthorhombic (top, green diamonds) and relaxed (bottom, black circles) NCs with experimental splittings.\cite{Rossi2020-JCP, Han2022, Yin2017, Nestoklon2018}  (f) Electron density plots for the HOMO (bottom) and LUMO (top) states. }
    \label{fig:figOptical}
\end{figure*}

We can evaluate the success of this method by comparing the computed optical gaps to a wide range of experimental results. In Fig. \ref{fig:figOptical}(a) we show the lowest excitonic states (dark and bright) of the relaxed NCs across a range of sizes. The excitation energies for the cubic and orthorhombic structures are plotted in Figure S6, and show a strong agreement with experimental PL measurements\cite{Swarnkar2016,Guvenc2019,Shang2019,Qiao2021,Paul2022}. The relaxed structures show a stronger confinement effect, with the smallest NCs having  higher excitation energies than the other structures due to the effects of relaxation on the angles between lead halide octahedra. The smallest relaxed structures differ significantly from either the cubic or orthorhombic geometries, and this forces the electron and hole quasiparticle states further apart in energy, opening the optical gap. For the larger NCs, the effect is the opposite as the optical gaps fall somewhat below that of the other structures and experiments. This can be understood through the simple bonding and anti-bonding picture of the bulk lead halide perovskites band structure. In the bulk, the upper valence band consists of antibonding states between Pb-6s and I-5p orbitals, and the lower conduction band consists of antibonding states between Pb-6p and I-5p orbitals, dominated by the Pb-6p orbitals.\cite{Brivio2014,Yan2019,Kang2017}

In the NCs the hole and electron quasi-orbitals (shown in Figure \ref{fig:figOptical}(f)) maintain much of their bulk character. For the smallest relaxed NCs, the Pb-I bond lengths are at a maximum, decreasing their antibonding interaction and lowering the valence band energy. The Pb-Pb distance is also exceptionally long lessening their interaction and pushing the conduction band higher in energy. This distortion changes character once the NCs pass the critical threshold of 3-4~nm in length where octahedral tilting brings the Pb atoms closer together, and the decreased Pb-I-Pb bond angle somewhat lessens the antibonding interaction. This brings the valence band quasiparticle energies into line with those of the orthorhombic structures, but the decreased Pb-Pb distances still drive the conduction band to fall below that of the fully orthorhombic structures. 

While the relaxation has some impact on the overall excitation energies, Figure \ref{fig:figOptical}(d) shows that it has little to no impact on the splitting between dark and bright states. For the cubic, orthorhombic and relaxed structures studied, the ground state exciton remains dark up to $6$nm NCs, and the trend with increasing size shows that a positive dark-bright splitting is expected for all sizes of NCs. This is consistent with the recent calculations by Biffi et al.~\cite{Biffi2022} which considered \ce{MAPbI_3} NC with atomistic electronic theory while only allowing relaxation of the MA cations. We find that expanding the relaxation to the lead halide octahedral backbone does not result in a level inversion and does not support a strong Rashba effect in these materials. If either  relaxation, or the enforcement of an orthorhombic crystal structure caused a significant Rashba effect, then the dark-bright splitting would be qualitatively different from that of the cubic structures which always have inversion symmetry and thus no Rashba effect. 

A deeper understanding of the exciton fine structure can come from investigating the spin statistics of the lowest excitonic states. While the total spin $\hat{S}_{tot}^2=\left(\hat{\bm{S}}_e+\hat{\bm{S}}_h\right)^2$ need not be a good quantum number, the expectation value of the total spin will still be indicative of the degree of spin-singlet versus spin-triplet character. As shown in Figure S5, the total spin expectation value for the dark states is very close to 2, the value for a triplet state. The bright states have lower total spin, indicative of greater spin-singlet character. The implications of this can be understood through the expectation values of the exchange interaction for each of the states. Only the bright states, with their partial spin-singlet character, feel the effects of exchange. This spin structure is present in all the structures we consider, ensuring a dark excitonic ground state regardless of structural relaxation.

The crystal structure does, however, have a significant impact on the splitting among the bright states as seen in Figure \ref{fig:figOptical}(e). The cubic structures are not shown as the bright levels are always degenerate. Considering the orthorhombic structures, the bright-bright splitting decreases with increasing NC size, contrary to the predictions of a model where the Rashba effect is sufficiently strong to cause a bright-dark inversion.\cite{Sercel2019} What is observed is consistent with a simple crystal-field splitting that would approach the bulk at large NC sizes.\cite{Fu2017,Ramade2018} The relaxed crystal structures are where we would expect to see signatures of the Rashba effect emerge if it was present, as the ions are allowed to relax and could strongly break inversion symmetry. These signatures are not present, and the  complex behavior that we do see in the bright-bright splittings  is due to the NC size dependent structural transition discussed earlier. The smallest NCs have more symmetric structures and thus smaller bright-bright splittings. For the larger NCs the structures become nearly orthorhombic and thus the splittings resemble those of the orthorhombic NCs. Both of these splittings match well to experiments,\cite{Yin2017,Nestoklon2018,Tamarat2019,Rossi2020-JCP,Han2022} although the lack of data for extremely small NCs make the predictions of the relaxed structures difficult to verify. Additional measurements of the exciton FS in extremely small perovskite NCs could help resolve questions of the structure of these smallest clusters.

The polarization dependent emission spectra can also be calculated from our atomistic theory. These spectra are shown in Figure S7 for NCs of various crystal structures all with 3.8~nm edge lengths. In agreement with the effective mass models, we observe only one peak for the cubic structure, lying around 20~meV above a dark ground state. For the orthorhombic structure, the splittings due to lattice distortion are clearly recovered and the polarization of the lower two excitonic states is aligned with the orthorhombic crystal axes rather than the faces of the NC, also in agreement with the effective mass models.\cite{Han2022} The spectra for the relaxed structures show similarities to the orthorhombic structure, but the lower two excitonic states are close enough in energy that they may not be resolvable into separate peaks. The results for the relaxed structures add the additional structural complexity not considered in an effective mass model. Taken as a whole, the results of our model conclusively show that the inversion symmetry breaking in perovskite NCs is not sufficient to produce a bright ground state. 

\begin{figure}
    \centering
    \includegraphics[width=7cm]{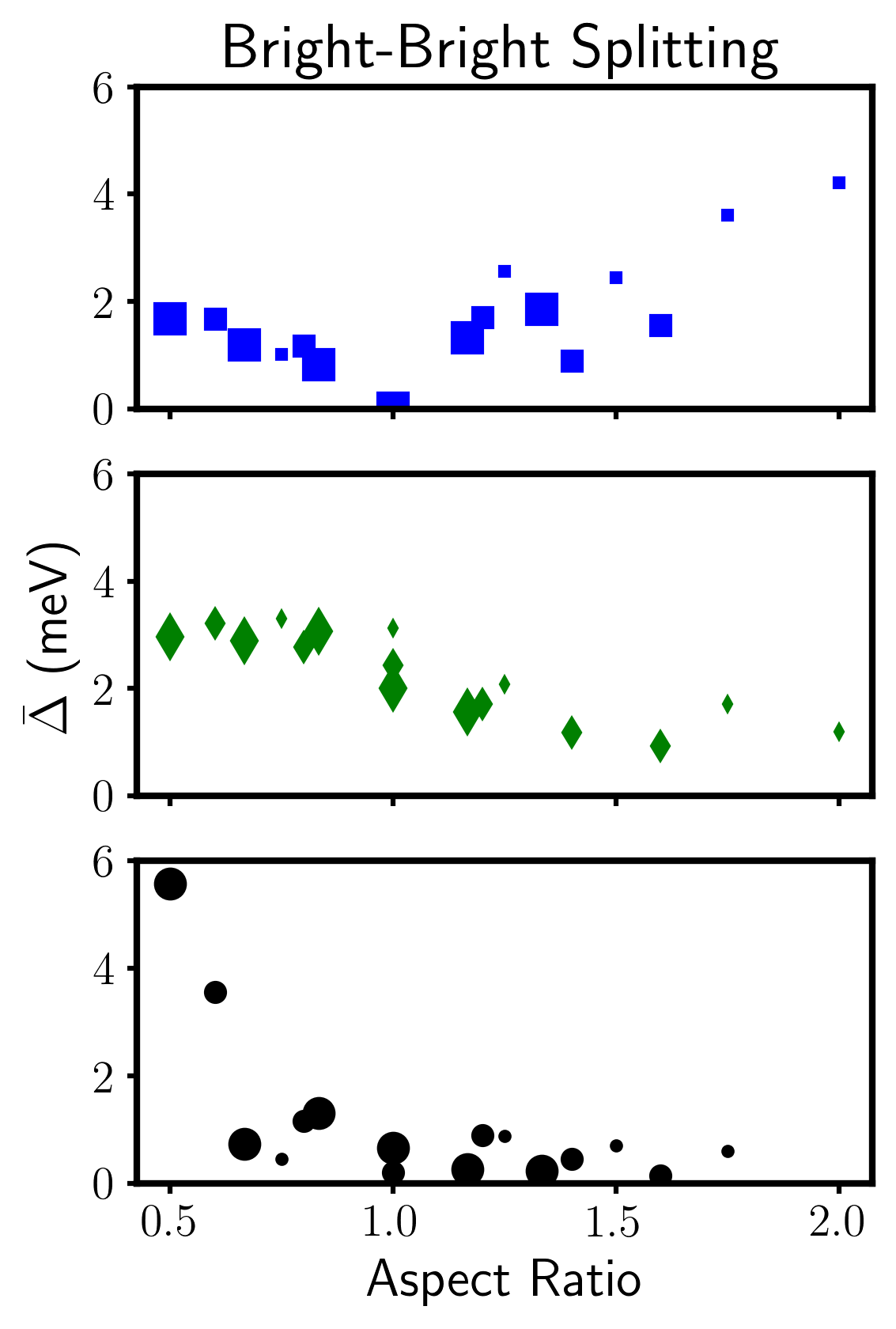}
    \caption{The standard deviation of the bright excitonic states for the cubic (top), orthorhombic (middle) and relaxed (bottom) crystal structures. The sizes of the symbols represent the sizes of the NCs along the x and y directions. The largest symbols correspond to $N=6$ and the smallest to $N=4$. }
    \label{fig:aniso}
\end{figure}

While it is easy to simulate a perfectly cubic NC, experiments tend to produce a distribution of NCs that differ from the perfect cubic geometry. This NC shape anisotropy is also known to impact the excitonic fine structure and may be implicated in the diversity of fine structure splittings observed experimentally.\cite{Tamarat2020} In Figure~\ref{fig:aniso} we consider the effect of shape anisotropy on NCs of cubic, orthorhombic and relaxed crystal structures. We generated a series of NCs consisting of $N\times N\times Z$ lead-halide octahedra where $N=4,5,6$ and $Z=3,\dots,8$. We define the aspect ratio as $Z/N$ and plot the standard deviation of the bright states for each of these NCs. For the NCs with cubic crystal structures the bright-bright splitting is zero for cube-shaped crystals, and either adding or removing layers from such a NCs causes a finite splitting. As the axis of shape anisotropy was chosen as the z-axis, the x- and y-polarized excitons remain degenerate. For an aspect ratio less than 1 the z-polarized exciton is split higher in energy. For an aspect ratio greater than 1 the z-polarized exciton is lower in energy than the x- or y-polarized ones. 

The NCs with an orthorhombic crystal structure show a significant degree of bright-bright splitting at all aspect ratios, consistent with effective mass theories.\cite{Han2022} The relaxed structures show a unique behaviour with significant bright-bright splittings at aspect ratios less than $1$, but a small and nearly constant splitting for aspect ratios larger than 1. This behaviour may result from surface relaxation effects that become more dominant for plate-like geometries. As single NC measurements remain extremely challenging, understanding the exact impacts of NC shape anisotropy is still an experimental challenge.

In conclusion, we calculate the exciton FS for lead-halide perovskite NCs using a fully atomistic theory to obtain relaxed NC crystal structures and the electronic states of these relaxed NCs. The structural relaxation reveals in atomistic detail previously predicted structural transitions, and the electronic theory reproduces experimental optical gaps with excellent agreement. This atomistic theory would be able to discern the causes and nature of a Rashba effect caused by collective inversion symmetry breaking if it was present in these systems. None of the signatures of a significant Rashba effect are found in this study. For all NCs studied the excitonic ground state is optically dark, and we conclude that it should remain so for all NC sizes. The explanation of the anomalous temperature dependence of radiative lifetimes in perovskite NCs must lie elsewhere and will be a subject of future investigation.

\section*{Acknowledgments}
This work was supported by the U.S. Department of Energy, Office of Science, Office of Basic Energy Sciences, Materials Sciences and Engineering Division, under Contract No. DE-461AC02-05-CH11231 within the Fundamentals of Semiconductor Nanowire Program (KCPY23).Methods used in this work were provided by the Center for Computational Study of Excited State Phenomena in Energy Materials (C2SEPEM), which is funded by the U.S. Department of Energy, Office of Science, Basic Energy Sciences, Materials Sciences and Engineering Division, via Contract No. DE-AC02-05CH11231 as part of the Computational Materials Sciences program.  This research used resources of the National Energy Research Scientific Computing Center (NERSC), a U.S. Department of Energy Office of Science User Facility. D.W is grateful for fruitful discussion with John Philbin and  Ming Chen on the implementation of spin-orbit coupling. E.R. is grateful to Sasha Efros for fruitful discussions.  We are thankful for useful conversations with Timothy Berkelbach. 

\section*{Supporting Information}
Additional details on the structural and electronic calculations, including the parameterization of pseudopotentials can be found below.

An online repository of the codes and data used in this paper can be accessed at \verb|https://github.com/dgweinberg/perovNCs| 

\section*{References}
\bibliographystyle{aipnum4-2}
\bibliography{bibliography}

\clearpage

\end{document}


\title{Supporting information: Size-dependent lattice symmetry breaking determines the exciton fine structure of perovskite nanocrystals}

\author{Daniel Weinberg, Yoonjae Park, David Limmer, Eran Rabani }

\maketitle

\section{Nanocrystal Structures}

\subsection{Unrelaxed Nanocrystal Structures}

The cubic and orthorhombic nanocrystal (NC) structures were generated
to correspond with the literature bulk crystal structures\cite{Sutton2018}.
A large slab of the bulk material was generated and the NC was and
cut from the bulk. The cut was made along the 100 facets of the cubic
structures and the 110 and 001 facets in the orthorhombic structures.
The cuts were made to ensure that all lead-iodide octahedra were complete
and the surface termination was only cesium and iodine. The Cs atoms at the corners of the NCs were found to be particularly unstable when relaxed, so these were removed for all structures. The cubic NCs considered in Figure 3 are listed in Table 
\ref{tab:CubicStructTable} and follow the general formula that for a cubic NC with $N$ Pb atoms along the each axis, there will be $(N +1)^3-8$ Cs atoms, $N^3$ Pb atoms, and $3(N^3+N^2)$ I atoms. The additional cuboidal NCs considered in Figure 4 have $N$ Pb atoms along the $x$- and $y$-axes and $Z$ Pb atoms along the $z$-axis. The configurations are listed in Table \ref{tab:CuboidStructTable} and follow the general formula that there will be $Z(N+1)^2-8$ Cs atoms, $ZN^2$ Pb atoms, and $3ZN^2+N^2+2ZN$ I atoms.

\subsection{Relaxed Nanocrystal Structures}

For each size of NC, we take the unrelaxed orthorhombic structure as an initial configuration.  However, since the corresponding structure doesn't satisfy the condition of charge neutrality, different partial charge is assigned for surface Cs atoms to stabilize the NC structure and reduce the effect from surface boundaries of NC. Using the force field parameters adopted from Ref.\cite{Bischak2020} which are parameterized to reproduce the energy difference between different crystal structures, the partial charge of surface Cs atoms, $q_{\textrm{surf}}$, is defined as 
\begin{equation}
q_{\textrm{surf}} = - \frac{N_\textrm{Pb} \, q_{\textrm{Pb}} \, + \, N_\textrm{I} \, q_{\textrm{I}} \, + \, N_\textrm{Cs} \, q_{\textrm{Cs}}}{N_{\textrm{surf}}}
\end{equation}
where $q_{\alpha}$ is a partial charge of atom $\alpha$ listed in Table S1 and $N_{\alpha}$ is the number of $\alpha$ atoms in each NC with $\alpha \in \{\textrm{Pb}, \, \textrm{I}, \, \textrm{Cs} \}$. The subscript $\textrm{surf}$ is used to refer to surface Cs atoms whereas the subscript $\textrm{Cs}$ indicates the core Cs atoms.    
With the modified initial configuration, the structure of each NC size is minimized using conjugate gradient algorithm based on the pairwise interaction $u_{ij}$ between atom $i$ and $j$ described by Lennard-Jones potential with Coulombic interaction 
\begin{equation}
u_{ij}(r) = \frac{q_i q_j}{4\pi \varepsilon_0 r} + 4 \varepsilon_{ij}
\left[ \left( \frac{\sigma_{ij}}{r} \right)^{12} - \left( \frac{\sigma_{ij}}{r} \right)^{6} \right] \, , \ \ r < r_c
\end{equation} 
where $i,j \in \{ \textrm{Pb}, \, \textrm{I}, \, \textrm{Cs} , \, \textrm{surf} \}$, $\varepsilon_0$ is the vacuum electric permittivity, and the cutoff distance $r_c$ is set to be the maximum length of each NC among three different axis times $\sqrt{3}$ to take care of the fact that periodic boundary condition cannot be applied for NC. Lennard-Jones parameters $\varepsilon$ and $\sigma$ of each atom are listed in Table S1. Where parameters not listed can be derived using following combining rules $\varepsilon_{ij} = \sqrt{\varepsilon_{i}\varepsilon_{j}}$ and $\sigma_{ij} = (\sigma_i + \sigma_j)/2$. Minimizations are performed using the LAMMPS package \cite{Thompson2022}.  

A DFT structural minimization was also performed on a 1.26~nm NC using the PBE exchange correlation functional. The electronic calculation used a kinetic energy cutoff of 65 Rydbergs and a charge density cutoff of 530 Rydbergs. The structure was optimized to a force threshold of less than $10^-4$ atomic units using Quantum Espresso\cite{Giannozzi2009,Giannozzi2017,Giannozzi2020}

\section{Single Particle Electronic States}

\subsection{Pseudopotential Method}

The ground state electronic calculations for a single phase were carried
out using the semi-empirical pseudopotential method~\cite{Wang1994,Wang1996,Rabani1999b,Williamson2000}, which assigns
to each atom in the NC an effective potential derived from bulk band
structures. The electronic Hamiltonian is a single electron operator
\begin{equation}
\hat{H}=\hat{T}+\sum_{\alpha}^{\text{atoms}}\left[\hat{V}_{loc}^{\alpha}+\hat{V}_{nonloc}^{\alpha}+\hat{V}_{SO}^{\alpha}\right]
\end{equation}
where $\hat{T}$ is the kinetic energy operator, $\hat{V}_{loc}^{\alpha}$
is the local part of the pseudopotential around atom $\alpha$, $\hat{V}_{nonloc}^{\alpha}$
describes angular momentum-dependent corrections to the local pseudopotential
around atom $\alpha$, and $\hat{V}_{SO}^{\alpha}$ describes the
spin orbit coupling around atom $\alpha$. The local part of the potential
is defined in by a reciprocal space function
\begin{equation}
\tilde{v}_{loc}^{\alpha}\left(q\right)=a_{0}^{\alpha}\frac{q-a_{1}^{\alpha}}{a_{2}^{\alpha}\exp\left(a_{3}^{\alpha}q^{2}\right)-1}
\end{equation}
where $q$ is the reciprocal coordinate, and the parameters $a_{0}^{\alpha}\dots\,a_{3}^{\alpha}$
are fit based on the atom $\alpha$. The potential is defined in terms
of the position-space counterpart of $\tilde{v}_{loc}^{\alpha}\left(q\right)$,
which we call $v_{loc}^{\alpha}\left(r\right)$. The local part of
the potential is given by 
\begin{equation}
\hat{V}_{loc}^{\alpha}=v_{loc}^{\alpha}\left(\left|\hat{\bm{r}}-\bm{R}_{\alpha}\right|\right)
\end{equation}
where $\hat{\bm{r}}$ is the position operator, $\bm{R}_{\alpha}$
is the position of atom $\alpha$. The angular momentum-dependent
part of the pseudopotential gives a correction to the local part of
the pseudopotential for the electrons in p-type orbitals. 
\begin{equation}
\hat{V}_{nonloc}^{\alpha}=\delta v_{l=1}^{\alpha}\left(\left|\hat{\bm{r}}-\bm{R}_{\alpha}\right|\right)\hat{P}_{l=1}^{\alpha}=\left[a_{4}^{\alpha}\exp\left(-\left|\hat{\bm{r}}-\bm{R}_{\alpha}\right|^{2}\right)+a_{5}^{\alpha}\exp\left(-\left(\left|\hat{\bm{r}}-\bm{R}_{\alpha}\right|-\rho\right)^{2}\right)\right]\hat{P}_{l=1}^{\alpha}
\end{equation}
where $\hat{P}_{l=1}^{\alpha}$ is the projector onto the $l=1$ angular
momentum subspace around atom $\alpha$, $\rho$ is a shift of 1.5
Bohr, and the $a_{4}^{\alpha}$ and $a_{5}^{\alpha}$ parameters are
fit based on atom $\alpha$. The spin-orbit coupling acts only on
the p-type orbitals as well and has the form
\begin{equation}
\hat{V}_{SO}^{\alpha}=v_{SO}^{\alpha}\left(\left|\hat{\bm{r}}-\bm{R}_{\alpha}\right|\right)\hat{\bm{L}}^{\alpha}\cdot\hat{\bm{S}}\hat{P}_{l=1}^{\alpha}=a_{6}^{\alpha}\exp\left(\frac{-\left|\hat{\bm{r}}-\bm{R}_{\alpha}\right|^{2}}{w^2}\right)\hat{\bm{L}}^{\alpha}\cdot\hat{\bm{S}}\hat{P}_{l=1}^{\alpha}
\end{equation}
where $\hat{\bm{L}}^{\alpha}$ is the vector of electron orbital angular momentum
operators around atom $\alpha$, $\hat{\bm{S}}$ is the vector of electron
spin operators, $w$ is a width of 0.7 Bohr and $a_{6}^{\alpha}$
is fit based on atom $\alpha$. The total potential can be rewritten
as three separate, spherically symmetric potentials felt by $s$-type
(along with $d$-type and higher angular momentum) orbitals, $p_{\frac{1}{2}}$-type
orbitals and $p_{\frac{3}{2}}$-type orbitals. The $s$-type orbitals
feel the local potential only:
\begin{equation}
v_{s}\left(r\right)=v_{loc}^{\alpha}\left(r\right)
\end{equation}
The $p_{\frac{1}{2}}$-type orbitals feel a combination of the angular
momentum dependent potential and the spin-orbit potential 
\begin{equation}
v_{p_{\frac{1}{2}}}\left(r\right)=v_{loc}^{\alpha}\left(r\right)+\delta v_{l=1}^{\alpha}\left(r\right)-v_{SO}^{\alpha}\left(r\right)
\end{equation}
while the $p_{\frac{3}{2}}$-type orbitals feel a different combination
\begin{equation}
v_{p_{\frac{3}{2}}}\left(r\right)=v_{loc}^{\alpha}\left(r\right)+\delta v_{l=1}^{\alpha}\left(r\right)+\frac{1}{2}v_{SO}^{\alpha}\left(r\right)
\end{equation}

The single particle Hamiltonian was solved via the filter diagonalization
method~\cite{Wall1995,Toledo2002} on a real-space grid with a grid spacing of 0.5 Bohr. This
finer grid spacing was used to ensure sufficient convergence of the
non-local parts of the Hamiltonian. The non-local operators were implemented
via a modified Kleinman-Bylander representation\cite{Kleinman1982,Bloch1990,King-Smith1991}. On the order of a few hundred states in the energy range near the band gap were converged. 

\subsection{Parameterization of Pseudopotentials }

The pseudopotential parameters were fit for each atom in a particular crystal
structure in order to reproduce bulk band structures. For the perovskite system we are investigating here, we
generated best fit parameter sets for both the cubic and the orthorhombic
crystal structures. The best fit was determined by using the pseudopotential
Hamiltonian within a converged plane-wave basis to generate the bulk band structures of the respective
phases, with care taken to properly describe the non-local and spin-orbit interactions~\cite{Hybertsen1986, Wang1996, Wang1999, Bester2003}. These pseudopotential band structures were then compared to literature
GW band structures~\cite{Sutton2018}. 
In order to better compare with experimental results, the band gap
of the GW calculations for the cubic structure, which differed significantly
from measurements of the bulk band gap, were corrected by a static
shift of the valence band. The GW calculations of the orthorhombic
band structure agreed with experiment and were used without modification.
The valence band offsets between the cubic and orthorhombic phase
were calculated from DFT using Quantum Espresso~\cite{Giannozzi2009,Giannozzi2017}.

The seven pseudopotential parameters per atom were fit for each phase using
a Monte-Carlo fitting procedure where the objective function emphasized
the closeness of fit around the band gap as well as the effective
mass of the bands at the gaps. The parameter space was extensively
searched to find the best parameters. Initial fitting showed that
contribution from the Cs ion was nearly zero, consistent with the
understanding across lead halide perovskite materials that the valence
and conduction bands are composed mainly of lead and halide orbitals~\cite{Brivio2014}.
Thus the potential around the Cs atoms was set to zero and the fits
were further refined considering only the lead and iodine parameters.

The results of the fitting are shown in Figure \ref{fig:Fit-band-structures}. The orthorhombic band structure generated from our pseudopotentials has a hole effective mass of 0.320~$m_e$ and an electron effective mass of 0.386~$m_e$. The cubic band structure generated from our pseudopotentials has a hole effective mass of 0.289~$m_e$ and an electron effective mass of 0.309~$m_e$. These slightly overestimate the masses of the literature structures, leading to some small additional confinement effects in the NCs.

\subsection{Interpolation of Pseudopotentials for Relaxed Structures}

For relaxed nanocrystal structures, there is no bulk phase that perfectly
matches the nanocrystal structure. This requires that the pseudopotential
around each atom adapt to the local structure. We do this by linearly
interpolating between the pseudopotentials for the cubic and orthorhombic
structures. The relaxed nanocrystal Hamiltonian is then
\begin{equation}
\hat{H}=\hat{T}+\sum_{\alpha}^{\text{atoms}}\left(x_{\alpha}\right)\left[\hat{V}_{loc}^{\alpha,\text{ortho}}+\hat{V}_{nonloc}^{\alpha,\text{ortho}}+\hat{V}_{SO}^{\alpha,\text{ortho}}\right]+\left(1-x_{\alpha}\right)\left[\hat{V}_{loc}^{\alpha,\text{cubic}}+\hat{V}_{nonloc}^{\alpha,\text{cubic}}+\hat{V}_{SO}^{\alpha,\text{cubic}}\right]
\end{equation}
where $x_{\alpha}\in[0,1]$ denotes the extent of orthorhombic distortion.
For iodine atoms bonded to two lead atoms this is determined by the
Pb-I-Pb bond angle, $\theta_{\alpha}$. For the cubic phase that bond
is straight, $\theta=180^{\circ}$ while for the orthorhombic phase
there are two bond angles of $\theta_{\alpha}=150.8^{\circ}$ and
$\theta_{\alpha}=160.6^{\circ}$. Thus the orthorhombic distortion
was calculated as 
\begin{equation}
x_{\alpha}=\frac{180-\theta_{\alpha}}{180-106.6}
\end{equation}
where if the angle was less than 106.6 degrees the atom was assigned
a fully orthorhombic pseudopotential. Iodine atoms not bonded to two
lead atoms (dangling iodine at the surface) were assigned the cubic
pseudopotential. Lead atoms were assigned the average of the distortion
parameters of bonded iodine atoms. It is important to note that for
the local part of the pseudopotential, $\hat{V}_{loc}$, a linear
interpolation between the orthorhombic and cubic operators does not
mean a linear interpolation of the parameters listed in Table~\ref{tab:Best-pseudopotential-parameters}. 

The linear interpolation was deemed reasonable based on the closeness
of the pseudopotentials for the cubic and orthorhombic phases, as shown
in Figure~\ref{fig:Pseudopotentials-Plot}. To ensure that this interpolation
did not cause significant issues with with electronic structure, band
structures were calculated using unit cells with distortions between
those of the cubic and orthorhombic structures. The energies of the valence and conduction band edges from these calculations are shown in Figure~\ref{fig:bandInterp} as the structure is distorted from the cubic to the orthorhombic form. Because the orthorhombic
structure differs from the cubic structure only by slight rotations
of the lead-iodide octahedra it was simple to construct such structures
and calculate their band structures both using the interpolated pseudopotentials
and DFT, again within the Quantum Espresso~\cite{Giannozzi2009,Giannozzi2017} package. The DFT results
are known to significantly underestimate the band gap of the cubic
structure, but our pseudopotential method, being trained on the corrected
band structures, is able to match much better to experimental results
at the end-points. The interpolated pseudopotentials generate non-monotonic
trend in the band gap with overall a slight decrease in the band gap
over the sequence of structures from cubic to orthorhombic, while
the DFT shows a monotonic increase. However, the DFT increase seems
to stem only from the previously mentioned underestimation of the
band gap in the cubic phase. Without resorting to extremely computationally
costly techniques like GW, we are satisfied that the pseudopotentials
are able to smoothly interpolate between the cubic and orthorhombic
phases without issue.

\section{Excitonic States}

The excitonic states were calculated using the Bethe-Salpeter equation (BSE)
within the Tamm-Dancoff approximation, which writes the excitonic
states as linear combinations of non-interacting electron-hole pair
states~\cite{Rohlfing2000}. The $n$th excitonic state, $\left\vert \psi_{n}\right\rangle $,
is written as 
\begin{equation}
\left\vert \psi_{n}\right\rangle =\sum_{ai}c_{a,i}^{n}\left\vert a,i\right\rangle 
\end{equation}
where the indices $a,b,c,\dots$ refer to electron (unoccupied) states,
and the indices $i,j,k,\dots$ refer to hole (occupied) states, $\left\vert a,i\right\rangle $
refers to the non-interacting pair state, and the expansion coefficients, $c_{a,i}^{n}$,
are determined by the eigenvalue equation
\begin{equation}
\left(E_{n}-\Delta\varepsilon_{a,i}\right)c_{a,i}^{n}=\sum_{bj}\left(K_{ai;bj}^{d}+K_{ai;bj}^{x}\right)c_{b,j}^{n}
\end{equation}
which also determines the energy of the $n$th excitonic state $E_{n}$.
The electron hole interaction kernel, which describes the binding
of the independent electron-hole states into correlated excitonic
states, has two parts: the direct interaction $K^{d}$ describes the
coloumb attraction between the electron-hole pair, while the exchange
interaction $K^{x}$ controls the details of the excitonic spectrum,
crucially including the singlet triplet splitting under consideration
here. The direct interaction is calculated using a screened coulomb
interaction $W\left(\left|\bm{r}-\bm{r'}\right|\right)$ within a
static screening approximation with 
\begin{equation}
K_{ai;bj}^{d}=-\int d\bm{x}d\bm{x'}\phi_{a}^{*}\left(\bm{x}\right)\phi_{j}^{*}\left(\bm{x'}\right)W\left(\left|\bm{r}-\bm{r'}\right|\right)\phi_{i}\left(\bm{x'}\right)\phi_{b}\left(\bm{x}\right)
\end{equation}
The static dielectric constant used in the calculation is $\epsilon=6.1$\cite{Cho2019}.
The exchange interaction is calculated with the bare coulomb interaction
$v\left(\left|\bm{r}-\bm{r'}\right|\right)$ to be
\begin{equation}
K_{ai;bj}^{x}=\int d\bm{x}d\bm{x'}\phi_{a}^{*}\left(\bm{x}\right)\phi_{j}^{*}\left(\bm{x'}\right)v\left(\left|\bm{r}-\bm{r'}\right|\right)\phi_{b}\left(\bm{x'}\right)\phi_{i}\left(\bm{x}\right)
\end{equation}
The interaction kernel matrices are solved in the basis of band-edge
states. The binding energy of the $n$th excitonic state, $E^n_\text{B}$, is calculated as
\begin{equation}
    E^n_\text{B}=\left\langle \hat{K}^d+\hat{K}^x\right\rangle_n =\sum_{abij}\left(c^{n}_{a,i}\right)^{*}\left( K^d_{ai;bj}+K^x_{ai;bj}\right)c^{n}_{b,j} .
    \label{eqn:ExBind}
\end{equation}

Generally 60-80 electron states and a similar number of hole states were selected to form the basis for solving the BSE, which were sufficient to converge the excitonic fine structure for the low energy states under consideration here. 

The effect of the exciton spin on the exchange interaction is the key factor in energetically separating spin singlet and triplet states. The exciton spin operator is the sum of the electron and hole spin operators,
\begin{equation}
    \hat{\bm{S}}_\textrm{tot}=\hat{\bm{S}}_\textrm{e}+\hat{\bm{S}}_\textrm{h}
\end{equation}
The total spin of an excitonic state can be calculated as,
\begin{align}
    \left\langle \psi_{n}\right|\hat{S}_{\textrm{tot}}^{2}\left|\psi_{n}\right\rangle &=\left\langle \psi_{n}\right|\hat{S}_{\textrm{e}}^{2}+\hat{S}_{h}^{2}+2\hat{\bm{S}}_{\textrm{e}}\cdot\hat{\bm{S}}_{\textrm{h}}\left|\psi_{n}\right\rangle \\
    &=\sum_{ai;bj}c_{bj}^{n}\left(c_{ai}^{n}\right)^{*}\left\langle a, i\right|\left(\hat{S}_{\textrm{e}+}^{2}+\hat{S}_{\textrm{h}}^{2}+2\hat{\bm{S}}_{\textrm{e}}\cdot\hat{\bm{S}}_{\textrm{h}}\right)\left|b,j\right\rangle \\
    &=\sum_{ai;bj}c_{bj}^{n}\left(c_{ai}^{n}\right)^{*}\left(\left\langle a\right|\hat{S}_{\textrm{e}}^{2}\left|b\right\rangle \delta_{ij}+\left\langle i\right|\hat{S}_{\textrm{h}}^{2}\left|j\right\rangle \delta_{ab}+2\left\langle a\right|\hat{\bm{S}}_{\textrm{e}}\left|b\right\rangle \cdot\left\langle i\right|\hat{\bm{S}}_{\textrm{h}}\left|j\right\rangle \right)\\    &=\sum_{ai;bj}c_{bj}^{n}\left(c_{ai}^{n}\right)^{*}\left(3/4\delta_{ab}\delta_{ij}+3/4\delta_{ij}\delta_{ab}+2\left\langle a\right|\hat{\bm{S}}_{\textrm{e}}\left|b\right\rangle \cdot\left\langle i\right|\hat{\bm{S}}_{\textrm{h}}\left|j\right\rangle \right)\\
    &=3/2+2\sum_{ai;bj}c_{bj}^{n}\left(c_{ai}^{n}\right)^{*}\left\langle a\right|\hat{\bm{S}}_{\textrm{e}}\left|b\right\rangle \cdot\left\langle i\right|\hat{\bm{S}}_{\textrm{h}}\left|j\right\rangle 
\end{align}
Importantly, the spin of the hole states must be treated as that of time-reversed electronic states, meaning that while the electron spin matrix elements are simply,
\begin{equation}
    \left\langle a\right|\hat{\bm{S}}_\textrm{e}\left|b\right\rangle =\frac{1}{2}\left\langle a\right\vert \hat{\bm{\sigma}}\left\vert b\right\rangle 
\end{equation}
the hole spin matrix elements are given by
\begin{align}
    \left\langle i\right|\hat{\bm{S}}_\textrm{h}\left|j\right\rangle &=\frac{1}{2}\left[\left\langle i\right\vert\hat{\Theta}^\dagger\right] \hat{\bm{\sigma}}\left[\hat{\Theta}\left\vert j\right\rangle \right]\\
    &=-\frac{1}{2}\left\langle j\right\vert \hat{\bm{\sigma}}\left\vert i\right\rangle 
\end{align}
 where $\hat{\bm{\sigma}}$ is the vector of Pauli spin matrices.

Simulated spectra were calculated from the oscillator strengths of the transition from each excitonic state to the ground state. 
\begin{equation}
f_{(x,y,z)} \left(E\right) = \sum_n \delta\left(E-E_n\right) \sum_{ai} c_{a,i}^{n}\left\vert \left\langle i\right\vert \hat{\mu}_{(x,y,z)}\left\vert a\right\rangle \right\vert^2
\end{equation}
where $c_{a,i}^{n}$ are the BSE expansion coefficients for electron state $\left\vert a\right\rangle$ and hole state $\left\vert i\right\rangle$ into excitonic state $n$ with energy $E_n$. The dipole operator $\hat{\mu}_{x,y,z}$ is broken down along the three principal axes as shown in color in Figure \ref{fig:Spectra}, and the average of the three polarizations is used for the total spectrum shown as the black line in Figure \ref{fig:Spectra}.

\FloatBarrier
\clearpage

\begin{table}
    \centering
    \begin{tabular}{ccc}
        \hline
        Configuration & $N$ &$L$ (nm)\\
        \hline 
         \ce{Cs_{19}Pb_{8}I_{36}}& 2 &1.26 \\
         \ce{Cs_{56}Pb_{27}I_{108}} & 3&1.89\\
         \ce{Cs_{117}Pb_{64}I_{240}} & 4&2.52 \\
         \ce{Cs_{208}Pb_{125}I_{450}} & 5&3.14 \\
         \ce{Cs_{335}Pb_{216}I_{756}} & 6 &3.77\\
         \ce{Cs_{504}Pb_{343}I_{1176}} & 7 &4.40\\
         \ce{Cs_{721}Pb_{512}I_{1728}} & 8 &5.03\\
         \ce{Cs_{992}Pb_{729}I_{2430}} & 9 &5.66\\
    \end{tabular}
    \caption{Details of the cubic NC configurations}
    \label{tab:CubicStructTable}
\end{table}

\begin{table}
    \centering
    \begin{tabular}{cccccc}
        \hline
        Configuration & $N$ & $Z$ & $L$ (nm) & $L_z$ (nm) & aspect ratio\\
        \hline
        \ce{Cs_{92}Pb_{48}I_{184}} & 4 & 3 &2.52 &1.89 &0.75\\
        \ce{Cs_{142}Pb_{80}I_{296}} & 4 & 5 &2.52 &3.14 &1.25\\
        \ce{Cs_{167}Pb_{96}I_{352}} & 4 & 6 &2.52 &3.77 &1.5\\
        \ce{Cs_{192}Pb_{112}I_{408}} & 4 & 7 &2.52 &4.40 &1.5\\
        \ce{Cs_{217}Pb_{128}I_{464}} & 4 & 8 &2.52 &5.03 &2.0\\
        \ce{Cs_{136}Pb_{75}I_{280}} & 5 & 3 &3.14 &1.89 &0.6\\
        \ce{Cs_{172}Pb_{100}I_{365}} & 5 & 4 &3.14 &2.52 &0.8\\
        \ce{Cs_{244}Pb_{150}I_{535}} & 5 & 6 &3.14 &3.77 &1.2\\
        \ce{Cs_{280}Pb_{175}I_{620}} & 5 & 7 &3.14 &4.40 &1.4\\
        \ce{Cs_{316}Pb_{200}I_{705}} & 5 & 8 &3.14 &5.03 &1.6\\
        \ce{Cs_{188}Pb_{108}I_{396}} & 6 & 3 &3.77 &1.89 &0.5\\
        \ce{Cs_{237}Pb_{144}I_{516}} & 6 & 4 &3.77 &2.52 &0.67\\
        \ce{Cs_{286}Pb_{180}I_{636}} & 6 & 5 &3.77 &3.14 &0.83\\
        \ce{Cs_{284}Pb_{252}I_{876}} & 6 & 7 &3.77 &4.40 &1.17\\
        \ce{Cs_{433}Pb_{288}I_{996}} & 6 & 8 &3.77 &5.03 &1.33\\
    \end{tabular}
    \caption{Details of the cubiodal NC configuration}
    \label{tab:CuboidStructTable}
\end{table}

\begin{figure}
\begin{centering}
\includegraphics[width=240pt]{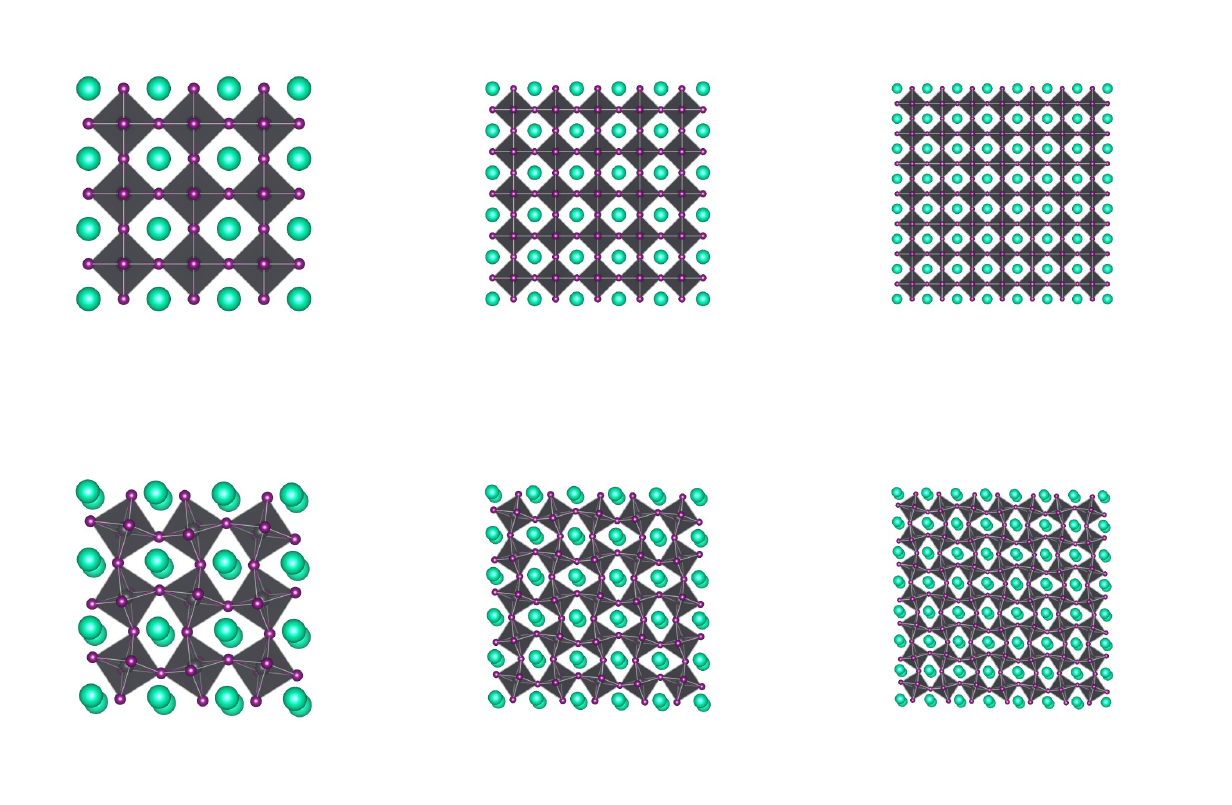}
\caption{Several sizes of cubic (top) and orthorhombic (bottom) NCs}
\end{centering}

\end{figure}

\begin{table}
\begin{centering}
\begin{tabular}{|c|c|c|c|c|}
\hline 
 & Pb & I & Cs & Cs (Surface)\tabularnewline
\hline 
\hline 
$q$ (e) & 0.85 & -0.57 & 0.86 & $q_{\text{surf}}$\tabularnewline
\hline 
$\sigma$ (Angstrom) & 3.210 & 4.014 & 3.584 & 3.584\tabularnewline
\hline 
$\varepsilon$ (eV) & 0.001086 & 0.06389 & 0.07728 & 0.07728\tabularnewline
\hline 
\end{tabular}\caption{Force field parameters from Ref. \cite{Bischak2020}}
\par\end{centering}
\end{table}

\begin{figure}
\begin{centering}
\includegraphics[width=0.45\columnwidth]{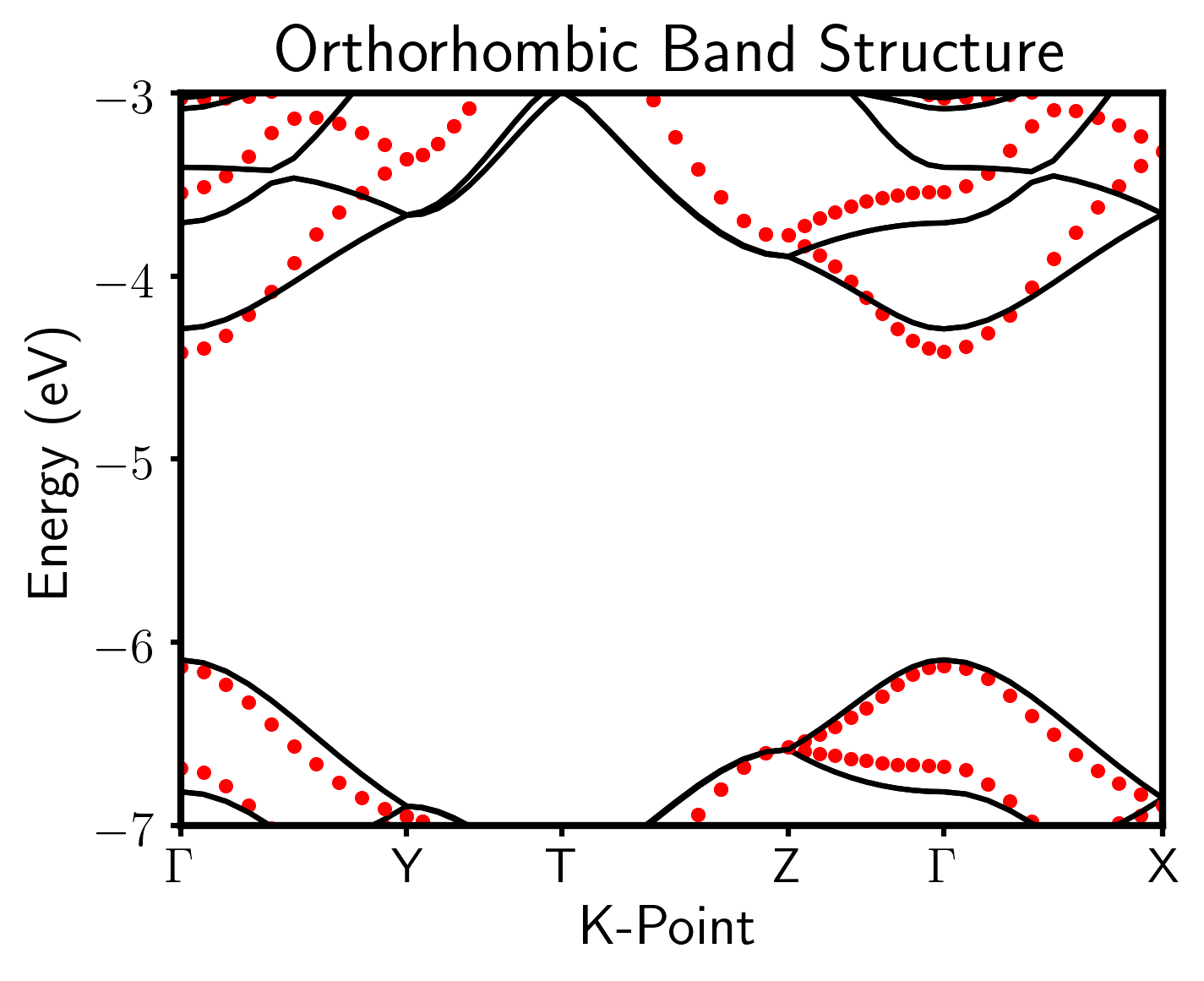}\includegraphics[width=0.45\columnwidth]{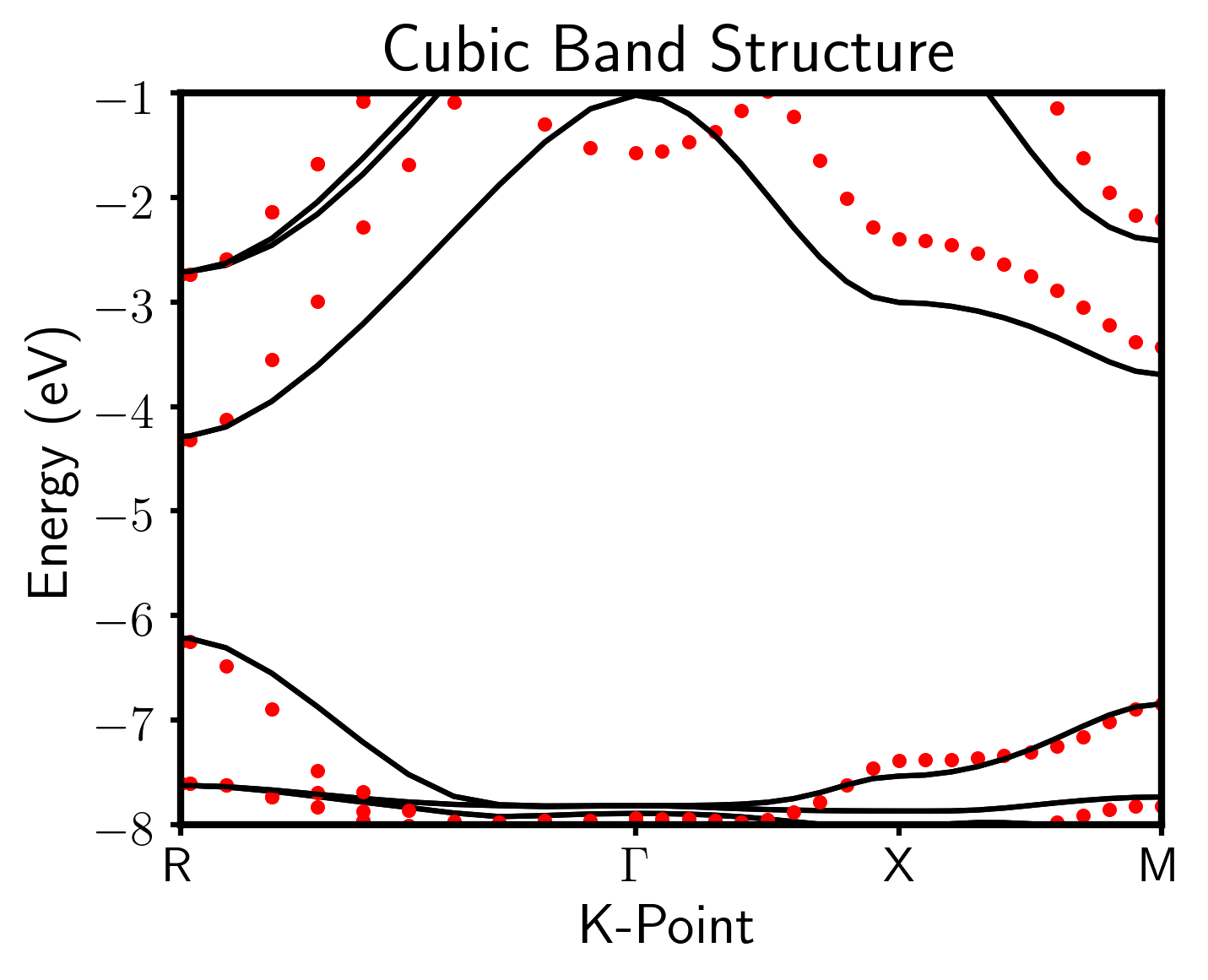}
\par\end{centering}
\caption{\label{fig:Fit-band-structures}The band structures generated by the
pseudopotential method (black lines) and the literature
band structures (red points). }
\end{figure}

\begin{table}
\begin{centering}
\begin{tabular}{|c|c|c|}
\hline 
Iodine  & Orthorhombic & Cubic\tabularnewline
\hline 
\hline 
$a_{0}$ & 117.89162750 & 113.75644925\tabularnewline
\hline 
$a_{1}$ & 2.12591587 & 2.14135468\tabularnewline
\hline 
$a_{2}$ & 2.91148249 & 2.84227140\tabularnewline
\hline 
$a_{3}$ & 0.58243028 & 0.56706414\tabularnewline
\hline 
$a_{4}$ & 0.16631532 & 0.04515637\tabularnewline
\hline 
$a_{5}$ & -0.02486412 & 0.00220169\tabularnewline
\hline 
$a_{6}$ & 2.23037552 & 3.03441874\tabularnewline
\hline 
\end{tabular}~~~~~ %
\begin{tabular}{|c|c|c|}
\hline 
Lead  & Orthorhombic & Cubic\tabularnewline
\hline 
\hline 
$a_{0}$ & 97.86083166 & 88.83030750\tabularnewline
\hline 
$a_{1}$ & 2.25710305 & 2.60529157\tabularnewline
\hline 
$a_{2}$ & 3.71951773 & 4.28692327\tabularnewline
\hline 
$a_{3}$ & 0.55872538 & 0.51950186\tabularnewline
\hline 
$a_{4}$ & 1.40412238 & 1.06405744\tabularnewline
\hline 
$a_{5}$ & -0.01566674 & -0.00808029\tabularnewline
\hline 
$a_{6}$ & 7.94721141 & 7.78990812\tabularnewline
\hline 
\end{tabular}
\par\end{centering}
\caption{\label{tab:Best-pseudopotential-parameters}Best pseudopotential parameters,
in Hartree atomic units.}

\end{table}

\begin{figure}
\begin{centering}
\includegraphics[width=0.45\columnwidth]{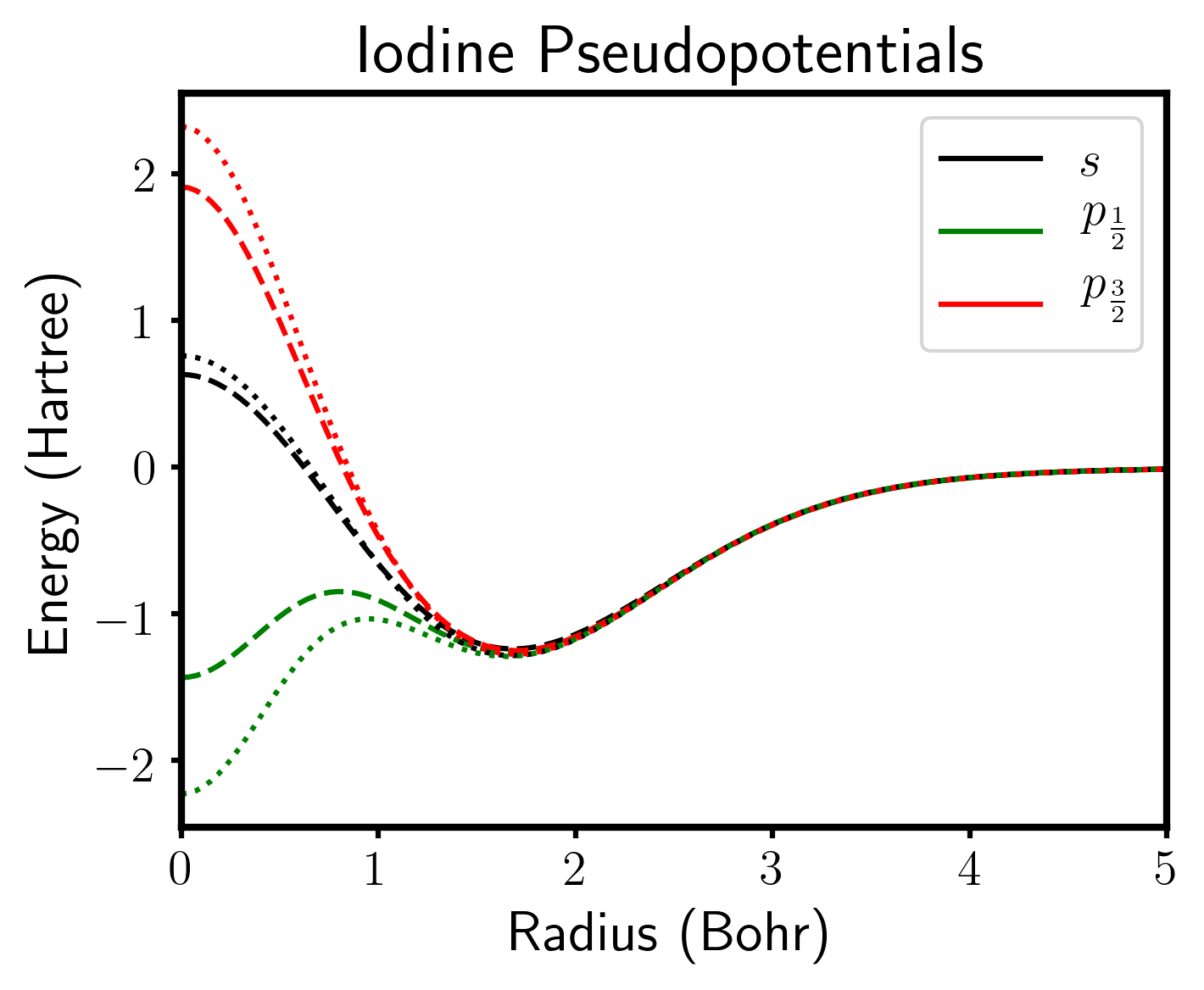}\includegraphics[width=0.45\columnwidth]{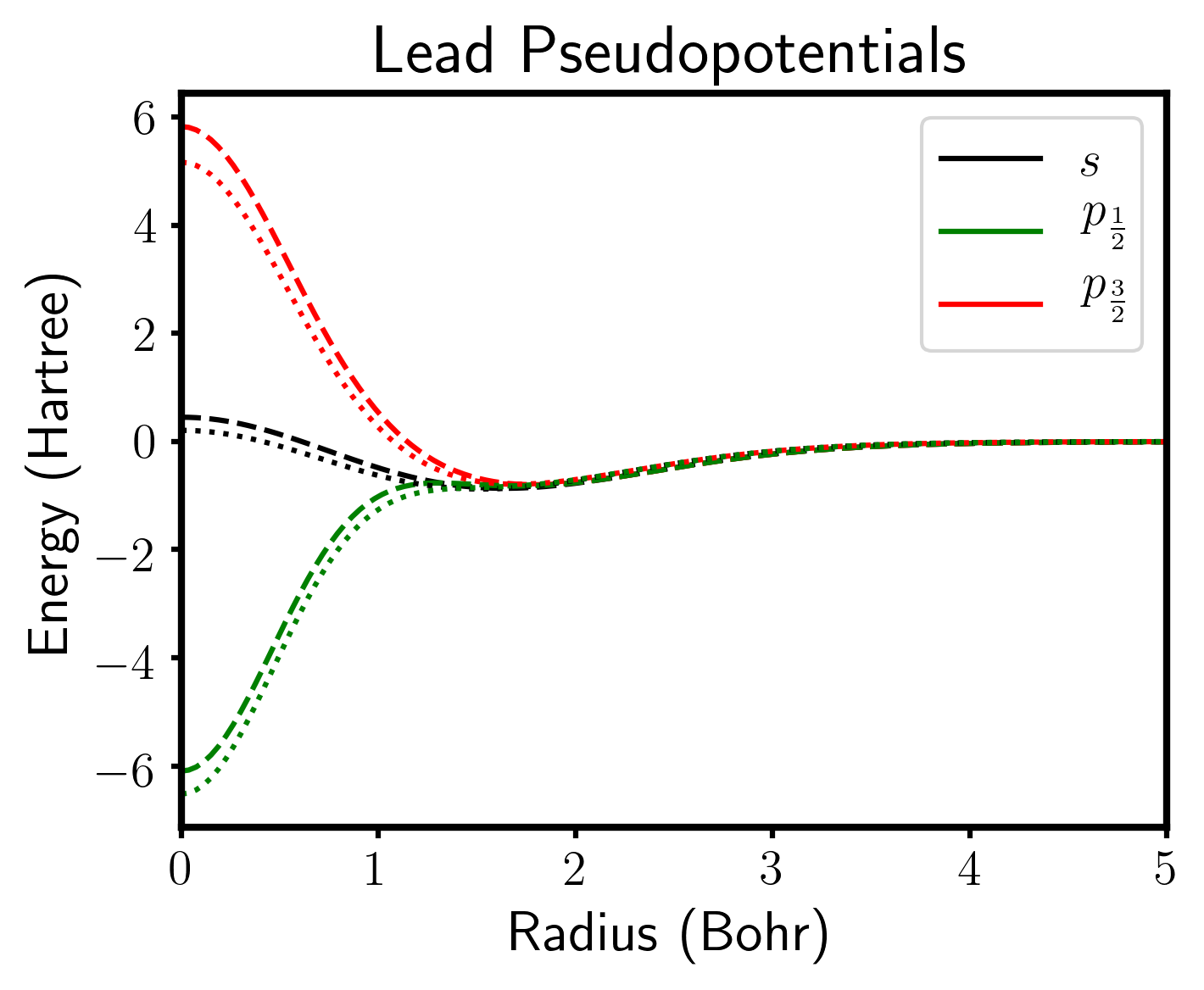}
\par\end{centering}
\caption{\label{fig:Pseudopotentials-Plot}Pseudopotentials for iodine and
lead atoms in the cubic (dotted lines) and orthorhombic (dashed lines). }
\end{figure}

\begin{figure}
\begin{centering}
\includegraphics[width=240pt]{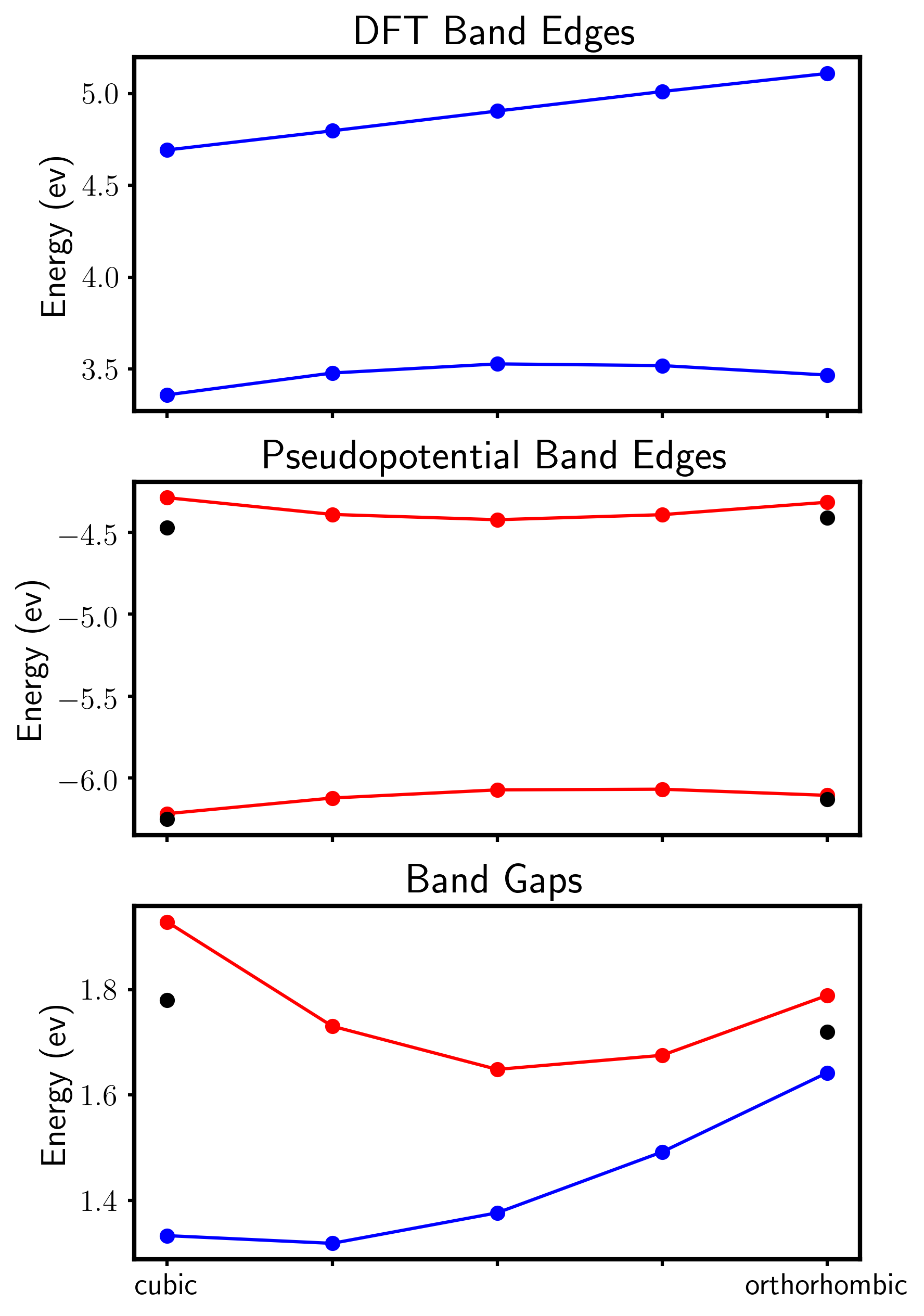}
\par\end{centering}
\caption{\label{fig:bandInterp}Comparison of DFT (blue) and pseudopotential (red) band edges in bulk
structures interpolated between the cubic and orthorhombic phases.}
\end{figure}

\begin{figure}
    \centering
    \includegraphics[width=0.9\columnwidth]{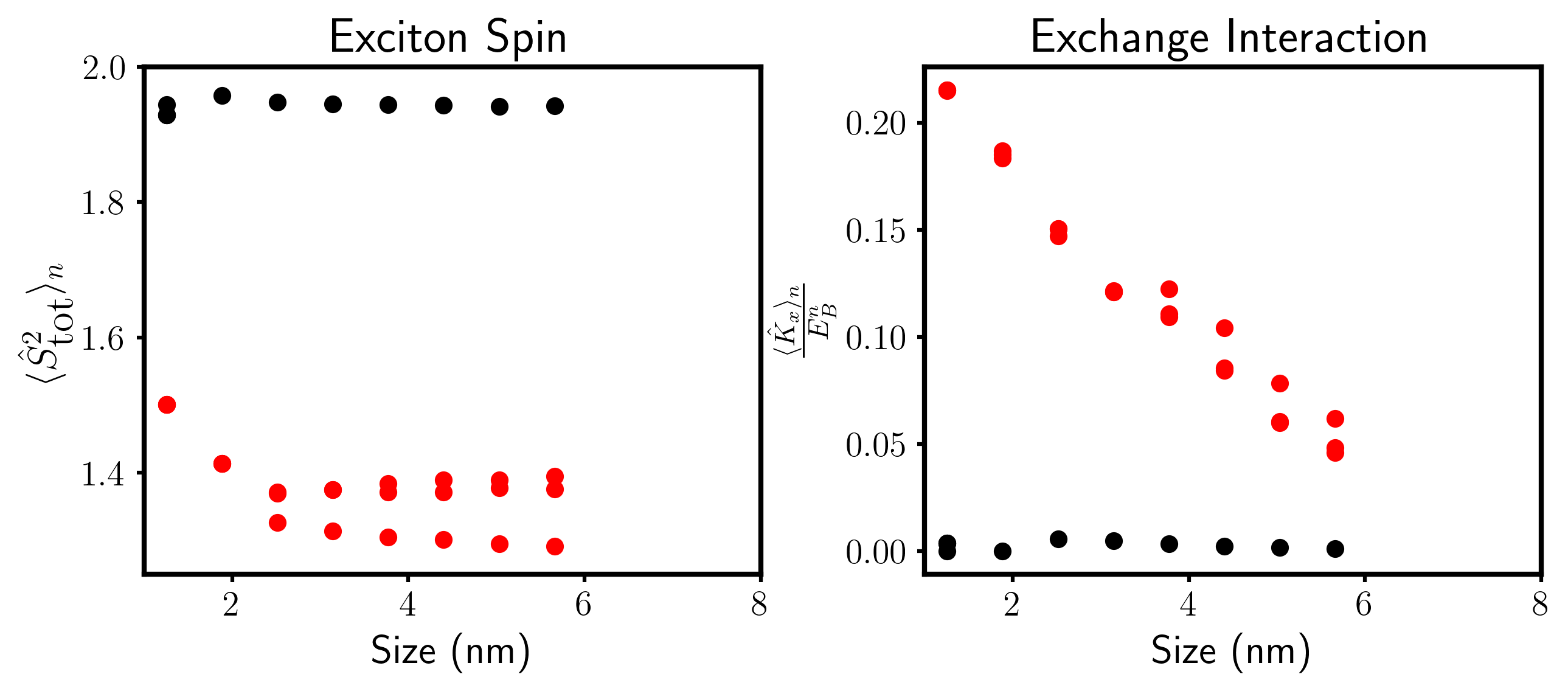}
    \caption{The total spin expectation value (left) and he exchange energy as a fraction of the total exciton binding energy (right) for the lowest dark (black) and bright (red) excitonic states as a function of size. }
    \label{fig:ExcitonSpins}
\end{figure}

\begin{figure}
    \centering
    \includegraphics[width=0.45\columnwidth]{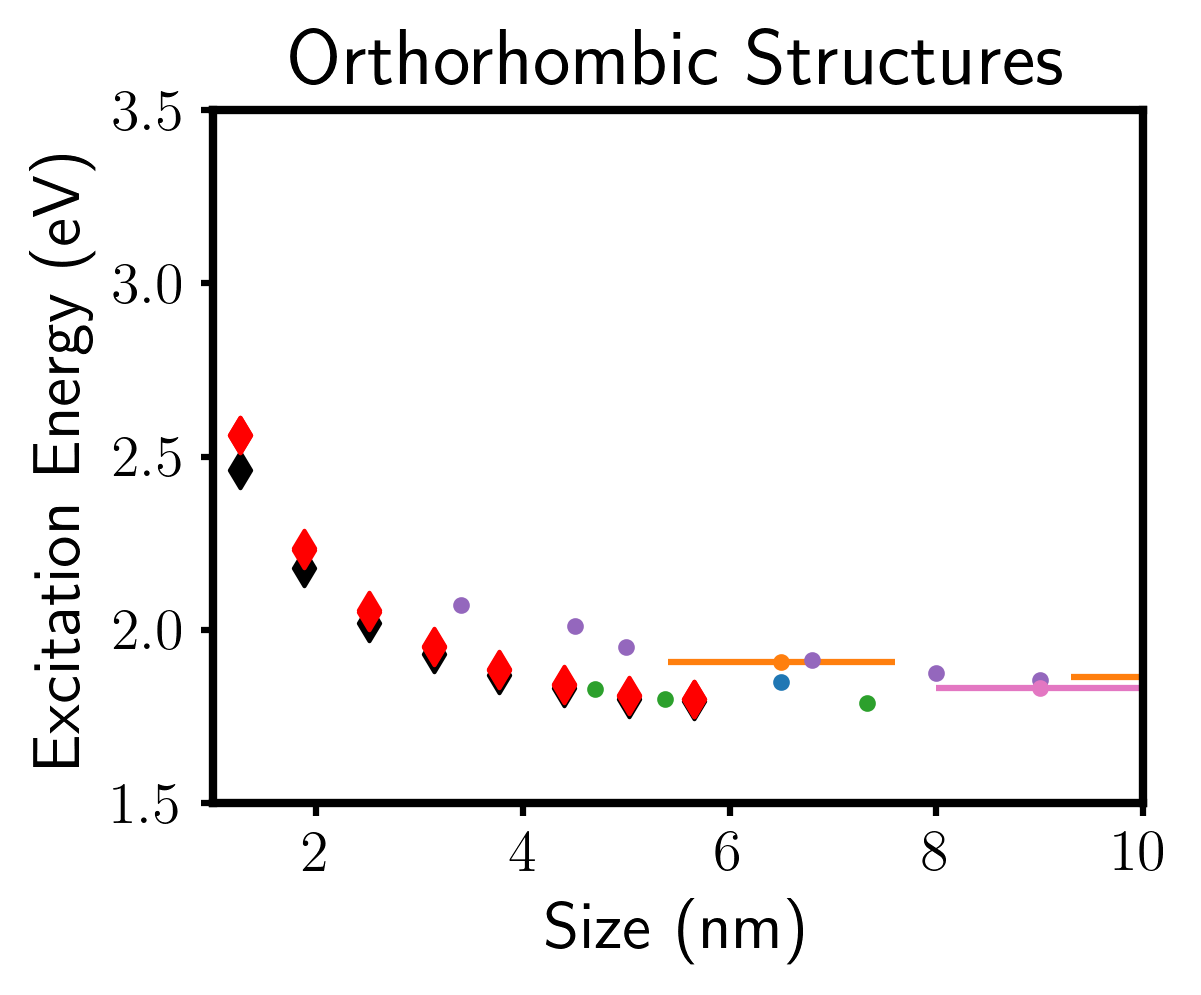}\includegraphics[width=0.45\columnwidth]{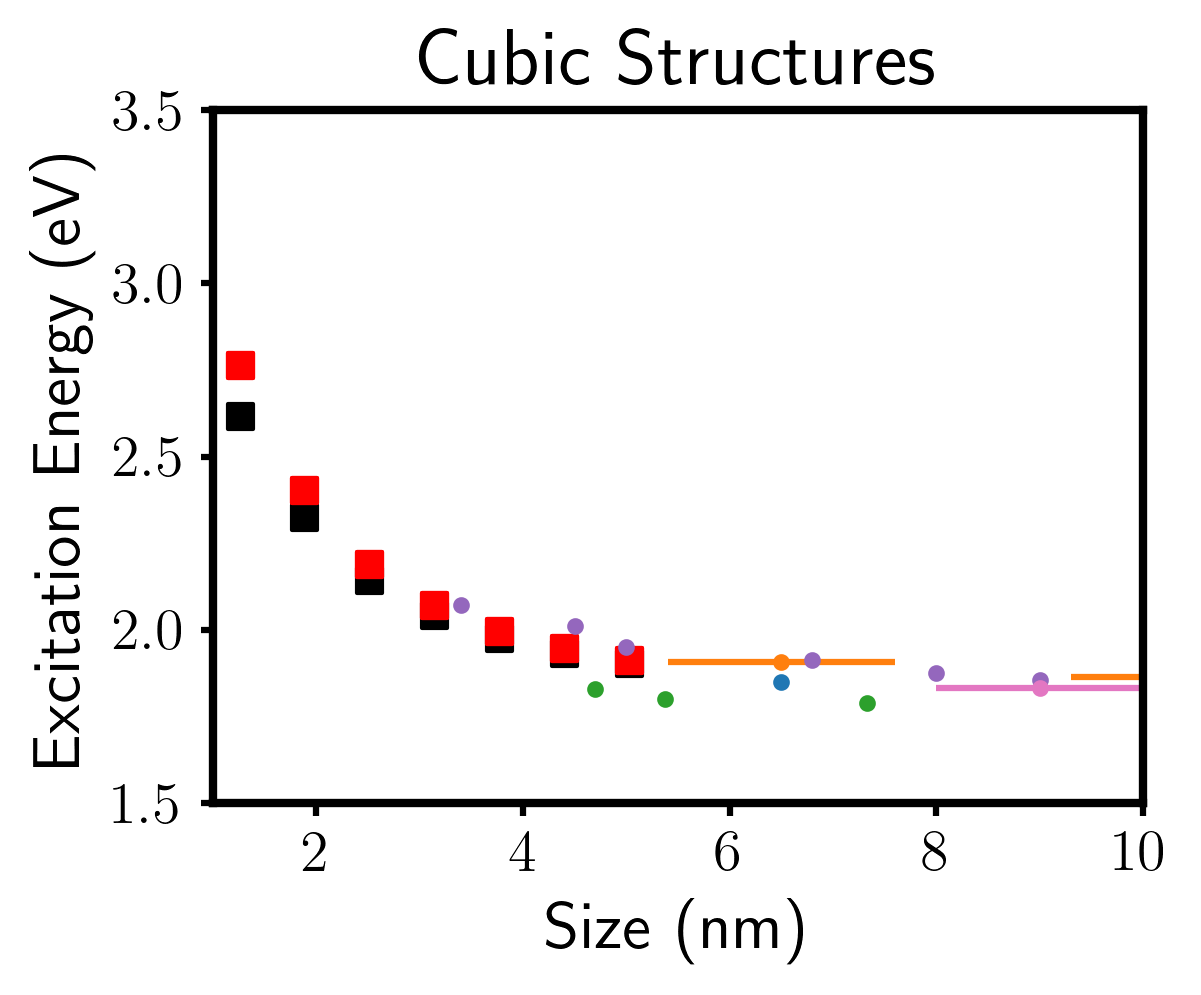}
    \caption{The lowest bright (red) and dark (black) excitonic states for NCs with orthorhombic  and cubic structures as a function of size along with experimental data \cite{Qiao2021,Shang2019, Yao2019, Swarnkar2016, Paul2022}.}
    \label{fig:OrthoCubicExcitons}
\end{figure}

\begin{figure}
    \centering
    \includegraphics[width=0.7\columnwidth]{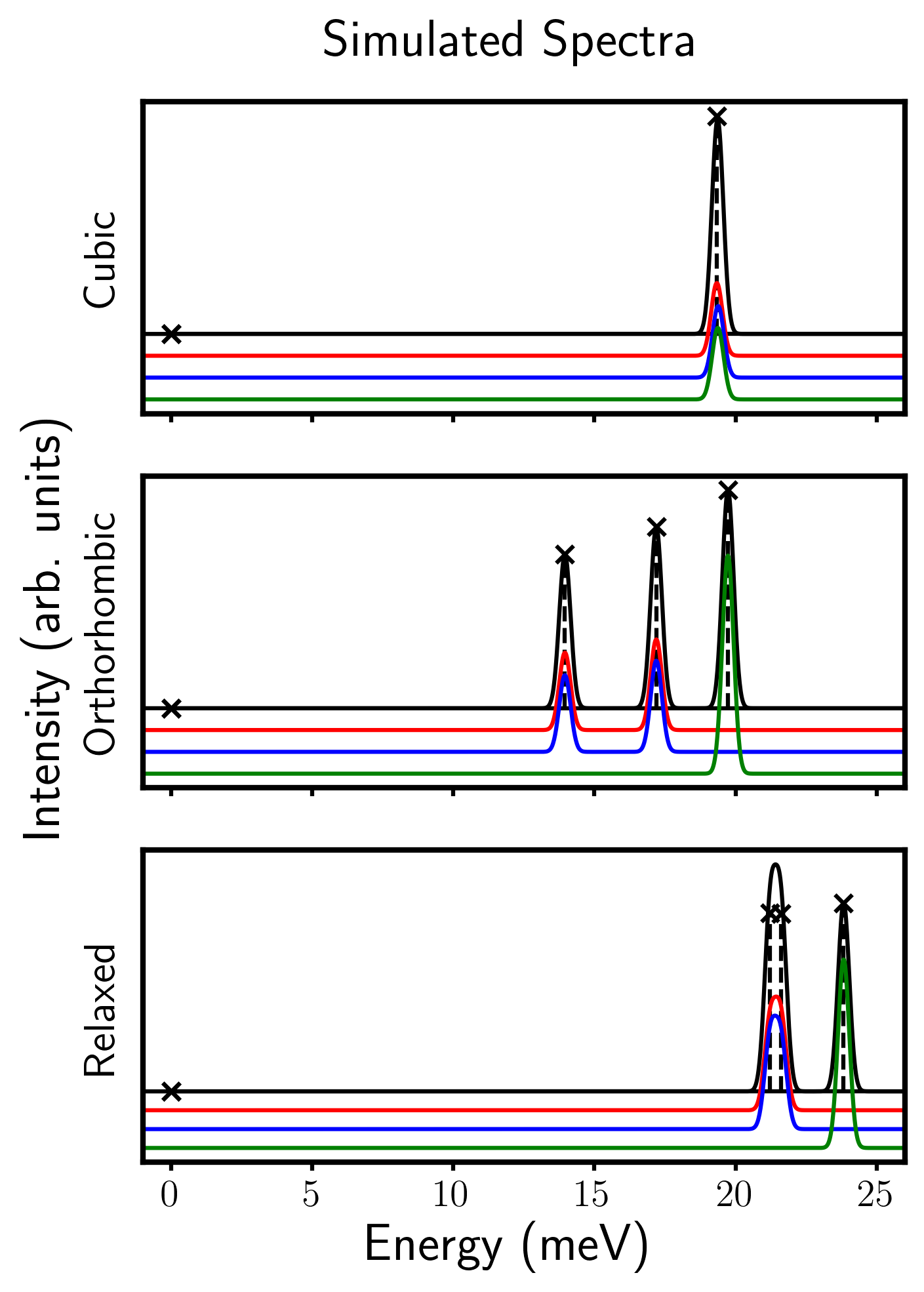}\\

    \caption{Simulated absorption spectrum for a 3.8~nm NC with various crystal structures with the lowest excitonic state set to the energy zero. The polarized spectra along the z (green), y (blue) and x (red) are also shown with and offset for clarity. }
    \label{fig:Spectra}
\end{figure}

\begin{table}
    \centering
    \begin{tabular}{ccc}
        \hline
        NC Size (nm) & Excitation Energy (eV) & Ref.\\
        \hline
         $6.5$ &$1.849$ &\cite{Qiao2021} \\
         $6.5\pm1.1$&$1.907$ &\cite{Shang2019}\\
         $10.8\pm1.5$&$1.864$ &\cite{Shang2019}\\
         $4.70$&$1.83$&\cite{Yao2019}\\
         $5.38$&$1.80$&\cite{Yao2019}\\
         $7.33$&$1.79$&\cite{Yao2019}\\
         $3.4$&$2.07$&\cite{Swarnkar2016}\\
         $4.5$&$2.01$&\cite{Swarnkar2016}\\
         $5.0$&$1.95$&\cite{Swarnkar2016}\\
         $6.8$&$1.91$&\cite{Swarnkar2016}\\
         $8.0$&$1.88$&\cite{Swarnkar2016}\\
         $9.0$&$1.86$&\cite{Swarnkar2016}\\
         $9.0\pm1$&$1.83$&\cite{Paul2022}\\
    \end{tabular}
    \caption{Experimentally measured excitation energies}
    \label{tab:ExpExcitation}
\end{table}

\begin{table}
    \centering
    \begin{tabular}{ccc}
        \hline
        NC Size (nm) & Splitting (meV) & Ref.\\
        \hline
        $4.9\pm0.2$ & 1.61 &\cite{Han2022}\\
        $5.4\pm0.3$ & 1.4 &\cite{Han2022}\\
        $6.2\pm0.3$ & 1.25 &\cite{Han2022}\\
        $6.7\pm0.45$ & 1.20 &\cite{Han2022}\\
        $7.4\pm0.6$ & 0.85 &\cite{Han2022}\\
        $7.9\pm0.7$ & 0.70 &\cite{Han2022}\\
        $9.3$       & 0.40 &\cite{Yin2017}\\
        $4.$        & 7.5  &\cite{Rossi2020-JCP}\\
        $6.5$       & 2.   &\cite{Rossi2020-JCP}\\
        $10.$       & 0.40 &\cite{Nestoklon2018}\\
    \end{tabular}
    \caption{Experimentally measured bright-bright splittings}
    \label{tab:ExpBrightSplit}
\end{table}

\clearpage
\FloatBarrier
\printbibliography